\documentclass[a4paper,11pt,fleqn]{article}
\usepackage[utf8]{inputenc}
\usepackage{rotating}
\usepackage[T1]{fontenc}
\usepackage{graphicx}
\usepackage{amsmath}
\usepackage{amsfonts}
\usepackage{amssymb}
\usepackage{xcolor}
\usepackage[colorlinks,citecolor=blue,linkcolor=blue]{hyperref}
\usepackage{float}
\usepackage{multicol}
\usepackage{booktabs}
\usepackage[authoryear]{natbib}
\date{}

\usepackage{flexisym}
\usepackage[font={sf},labelfont=bf]{caption}
\definecolor{orRev}{HTML}{D35400}
\definecolor{grRev}{HTML}{008c00}

\begin{document}
\let\WriteBookmarks\relax
\def\floatpagepagefraction{1}
\def\textpagefraction{.001}

% 
% \shorttitle{Kwinking in B19$^\prime$ NiTi martensite}
% \shortauthors{Seiner et~al.}
% 
% % title
\title{\bfseries Kwinking as the plastic forming mechanism\\ of  B19$^\prime$ NiTi martensite}                      
 \author{Hanu\v{s} Seiner$^{*\dag}$, Petr Sedl\'{a}k$^\dag$, Miroslav Frost$^\dag$, Petr \v{S}ittner$^\ddag$}

\maketitle

\begin{center}
\baselineskip18pt
{\small $^\dag$ Institute of Thermomechanics, Czech Academy of Sciences, Dolej\v{s}kova 5, 182 00 Prague, Czech Republic

$^\ddag$  Institute of Physics, Czech Academy of Sciences, Na Slovance 1999/2, 182 21, Prague, Czech Republic}

$^*$Corresponding author: hseiner@it.cas.cz

\end{center}

\begin{abstract}
	Irreversible plastic forming of B19$^\prime$ martensite of the NiTi shape memory alloy is discussed within the framework of continuum mechanics. It is suggested that the main mechanism arises from coupling between {{}martensite reorientation and coordinated $[100](001)_{\rm M}$ dislocation slip}. A heuristic model is proposed, showing that the ${(20\bar{1})_{\rm M}}$ deformation-twin bands, commonly observed in experiments, can be interpreted as a combination of dislocation-mediated kink bands, appearing due to strong plastic anisotropy, and reversible twinning of martensite. {{} We introduce a term 'kwinking' for this combination of reversible twinning and irreversible plastic kinking.}  The model is subsequently formulated using the tools of nonlinear elasticity theory of martensite and crystal plasticity, {{} introducing 'kwink interfaces' as planar, kinematically compatible interfaces between two differently plastically slipped variants of martensite. It}
	is shown that the ${(20\bar{1})_{\rm M}}$ kwink bands may be understood as resultsing from energy minimization, and that their nucleation and growth and their pairing with $(100)_{\rm M}$ twins into specific patterns enables low-energy plastic forming of NiTi martensite. {{} We conclude that kwinking makes plastic deformation of B19$^\prime$ martensite in polycrystalline NiTi possible despite only one slip system being available.}
		
\end{abstract}

 {{\bfseries Keywords:} Shape memory alloy (SMA);
  	Martensitic phase transformation;
 	Plastic forming;
  	Pattern formation;
  	Anisotropic dislocation slip;
     Crystal plasticity}

\renewcommand{\arraystretch}{2}

\section{Introduction}

While the most important functional properties of the NiTi shape memory alloy originate from the reversible martensitic transition \citep{Otsuka,Miyazaki_1989,Wei_1998}, irreversible plastic forming of NiTi is the key issue for employing this alloy in technological applications. The relation between plastic forming and the thermo-mechanical performance of the NiTi shape memory alloy is currently an extensively studied topic \citep{Polatidis_2020,plas2,Gao,PMS}; experimental observations indicate that the forming process is immensely complex, possibly involving plastic straining both in the austenite phase and in B19$^\prime$ martensite.  The plastic deformation of the cubic austenite phase by dislocation slip is relatively well described in the literature (see e.g. \citep{Chowdhury_Scripta_2016,Ezaz_Acta_2013} for a first-principles analysis of all possible slip systems). In contrast, the processes taking place in monoclinic  martensite when strained beyond the recoverability limits are much less explored yet. As observed first by \cite{Nishida_1998} and later by \cite{Zhang}, such a straining typically leads to formation of $(20\bar{1})_{\rm M}$ \emph{plastic twins},  that are never found in thermally induced, self-accommodates martensitic microstructures in B19$^\prime$. Upon heating, these interfaces are inherited in the austenite phase as residual $(41\bar{1})_{\rm P}$ twins \citep{Ii}, where the subscripts ${\rm M}$ and ${\rm P}$ denote the lattice indices corresponding to the martensite and the parent phase, respectively. As discussed by \cite{Gao, pathways}, the $(20\bar{1})_{\rm M}$ plastic twins are not predictable by the classical mathematical theory of martensitic microstructures \citep{JMB1,Bhattacharya,ZanPit}, as they correspond to straining beyond the given Eriksen-Pitteri neighborhood (EPN,\citep{Bhattacharya}), i.e., beyond the set of martensitic lattices related to the reference lattice of austenite by a unique lattice correspondence resulting from the transformation path \citep{Zelazny_2011}. This agrees well to the fact that the $(20\bar{1})_{\rm M}$ twins are not 'erased' by the reverse transition but are, instead, transformed into deformation twins found in the cubic lattice of austenite. To this extent, the importance of the $(20\bar{1})_{\rm M}$ twins for the plastic forming process is quite well understood.

Nevertheless, the mechanism how the $(20\bar{1})_{\rm M}$ plastic twins arise in irreversibly strained B19$^\prime$ has not been sufficiently explained yet, and there are also  
 several features of these twins that call for a more detailed analysis. The most apparent one is that the $(20\bar{1})_{\rm M}$ twinning provides shearing in the  
$(010)_{\rm M}$ lattice plane, i.e. in the same plane as several other deformation mechanisms of B19$^\prime$, namely the $[100]_{\rm M}(001)_{\rm M}$ dislocation slip and the $(100)_{\rm M}$ and $(001)_{\rm M}$ transformation twins. As shown in \citep{Ezaz_MSEA_2020,Ezaz_Acta_2013}, all these mechanisms are energetically favorable over the $(20\bar{1})_{\rm M}$ twinning, and thus, a question arises under which specific conditions this twinning system is activated, and how does it compare to other mechanism acting on the same plane in terms of achievable strains and their relative energetic expensiveness.
Secondly, the $(20\bar{1})_{\rm M}$ twins are typically observed as a part of particular V-shaped microstructures in combination with $(100)_{\rm M}$ twins (\citep{Zhang,Molnarova_2020}, see also Figures \ref{evolution} and \ref{wedge}). This means that such microstructures must be somehow energetically beneficial, either leading to energy minimization via pattern formation, or enabling a low-energy mechanism of plastic deformation. 

\bigskip
In this paper, we propose a micro-mechanical model of plastic straining of NiTi martensite, and formulate this model within the framework of continuum mechanics, combining the mathematical theory of martensite and crystal plasticity.  The main assumption of the model is that the reversible $(100)_{\rm M}$ and $(001)_{\rm M}$ twinning (i.e. twinning within one EPN) appears simultaneously with highly anisotropic dislocation slip. We show that such a model gives predictions consistent with the experimental observations, leading to energy reduction through formation of $(20\bar{1})_M$-twin bands and their coupling with $(100)_M$ twinning. The most important outcome of the model is the identification of a new plastic straining mechanism that can be described as \emph{twinning-assisted formation of symmetric-tilt grain boundaries}, or \emph{slip-assisted formation of twinned microstructures}. This mechanism can arise in a shape memory alloy martensite when the high energy barrier between different EPNs is overcome by dislocation slip; the unified modelling framework allows us to understand it as a result of energy-minimization (or at least energy-reduction).  The model suggest that the plastic forming mechanism of B19$^\prime$ is strikingly similar to the deformation mechanisms of layered media \citep{Wadee,LPSO} or materials with highly anisotropic dislocation slip \citep{Inamura,Lei}, which explains why also the resulting V-shaped patterns in B19$^\prime$ clearly resemble the patterns observed in these materials.

The second aim of this paper is to explore the properties of the newly suggested mechanism and to discuss them in the context of the current knowledge of plastic forming of NiTi, both from the theoretical/modelling point of view and as far as the experimental observations are concerned. Regarding the former, it is worth noting that although the continuum-level description of twinning in martensite and single-crystal plastic slip were well developed in the literature, their combined applications for capturing irreversible straining of shape memory alloys have been just sparsely reported so far \citep{Chowdhury}, and always with assuming that the dislocation slip appears only in the austenite phase. Motivated by experimental observations by \cite{Simon_2010, Delville_2011, Pelton_2012}, initiation of plastic slip of the austenite phase at the austenite-martensite interfaces were discussed by \cite{Paran1,Paran2} using quite complex multi-scale models, and several others \citep{Mo_2022, Chaugule_2022, Hossain_2021}. The dominant appearance of the plastic slip in austenite was also assumed for explaining the transformation-plasticity coupling in superelastically strained wires by \cite{PMS,plas1}, where this concept was shown to predict correctly the massive plastic slip associated with the reverse transition, as it has been recently confirmed experimentally for thermal cycling \citep{Akamine_2023} {{} as well as mechanical cycling \citep{Sidharth_Scripta_2023}}. However, these approaches cannot be applied for the case reported by \cite{Zhang} and \cite{Molnarova_2020} and discussed in this paper, since the austenite phase is absent here, and plastic deformation of martensite must be considered instead. The recent experimental literature concerned with plastic forming of martensite studied on the single-crystal level reveals enormous strain localization into finely microstructured bands \citep{Polatidis_2020} and partial reversibility of the plastic twinning \citep{Chen_2022,Alarcon_MatDes_2023}. These observations do not clarify the mechanism of the $(20\bar{1})$ twinning, neither do they explain the formation of the V-shaped patterns, but they prove that there are some twinning-plasticity coupling phenomena that go beyond the validity of the simplified models based on partitioning between twinning and slip in individual grains based on their crystallographic orientation and the respective Schmidt factors (see \citep{BhattaPoly, Kimiecik_JMPS_2016, Stebner_JMPS_2013} for detailed experimental and theoretical treatment of the partitioning). Notice also that while the $[100]_{\rm M}(001)_{\rm M}$ slip has been identified as the possible easiest dislocation slip of the B19$^\prime$ lattice already by \cite{Kudoh_1985} (which has been confirmed by first principles calculations by \cite{Ezaz_Acta_2011}), no dedicated experimental analysis of plastic anisotropy of NiTi martensite can be found in the available literature so far. By the model developed in this paper, we show that the $[100]_{\rm M}(001)_{\rm M}$ slip is essential for formation of the observed patters, which confirms the considerations of \cite{Kudoh_1985} and  theoretical results of \cite{Ezaz_Acta_2011}.

On the other hand, there is a quite extensive literature reporting on both experimental research and numerical simulations of NiTi plastic forming on the macro-scale, i.e. polycrystal-scale level. The majority of the works focus on the evolution of the macroscopic response during cyclic loading in pseudelasticity and describe the effective behavior of the material using the approaches developed for transformation/twinning-induced plasticity (TRIP/TWIP) in steels, either applying the approaches based on phenomenological plasticity, e.g. \citep{Lagoudas_2004, Auricchio_2007, Hartl_2009, Zaki_2010, Wang_2017, Petrini_2020, Scalet_2021, Song_2023}, or those inspired by the crystal plasticity and using some scale transition technique, e.g. \citep{Manchiraju_2010, BhattaPoly, Yu_2015, Xiao_2018}. Only a few models of the latter type include plastic processes running in martensite when no transformation occurs, e.g. \citep{Wang_2008, Yu_2014, Dhala_2019, Xu_2021, Ju_2022}, and in many of them, the plastic deformation of the martensite phase is assumed to come mainly from the twinning deformation running in 11 twinning systems, i.e. contribution of the $[100](001)_{\rm M}$ dislocation slip system is neglected \citep{Yu_2014, Dhala_2019, Ju_2022}.

\section{Preliminaries}
\subsection{Notation, terminology, and lattice parameters}

For both the discussion of the experimental observations and the construction of the model, we restrict our considerations to one crystallographic plane, which is the $(010)_{\rm M}$ plane in martensite in the standard notation (used e.g. in \citep{Bhattacharya}). The reason is that the deformation mechanisms that were experimentally observed as those most active in plastic forming of NiTi B19$^\prime$ martensite \citep{Zhang, Molnarova_2020, Sittner_SMSE_2023_SMST} have the shearing direction lying in this plane and the invariant plane perpendicular to it. The notation we use for these deformation mechanisms is summarized in Table \ref{tabul} together with their crystallographic parameters. There is no experimental evidence showing how the material accommodates strains in the direction perpendicular to the given $(010)_{\rm M}$ plane. For the purpose of this work, we assume that there might be a compensation of strains between clusters of neighboring grains, having differently oriented $(010)_{\rm M}$ planes. It is worth noting that the $(010)_{\rm M}$ plane orientations in the martensitic polycrystal are not strictly determined by the texture: the material always has an additional degree of freedom, as different grains can transform (or reorient) into different variants of martensite, which may ease such a compensation. For example, in a strongly $(111)_{\rm P}$-textured wire (with small equiaxed grains and rotationally symmetric texture \citep{Bian_AppMatToday_2022}), which is one of the cases discussed in this work, there are three variants of martensite that have the $[010]_{\rm M}$ direction oriented perpendicular to the loading axis, and thus, three variants in which the plastic straining in the $(010)_{\rm M}$ plane can directly contribute to the tensile elongation of the wire.

All twinning systems listed in Table \ref{tabul} are compound twins (in the sense as defined in e.g. \citep{ZanPit}), which means that their twinning plane is a lattice plane and {{}{the shearing direction is a lattice direction}}. The shearing magnitudes provided in Table \ref{tabul} were calculated using the lattice parameters from \citep{Bhattacharya}, which are: $a=2.889$~\AA, $c=4.622$~\AA, and $\beta=96.8^\circ$ \citep{Bhattacharya}. We use these parameters for all calculations in this paper, and also for  visualizations of the microstructures in figures (with the exception of schematic sketches in Figures \ref{evolution}, \ref{zigzage} and \ref{cartoon}); this allows us to discuss the compatibility in the microstructures and the strains carried by the microstructures directly from the visualizations.   

\begin{table}
\caption{A list of the considered deformation mechanisms acting on a selected $(010)_{\rm M}$ plane in B19$^\prime$ martensite, and the used notation. The twin components and the shearing magnitudes were taken from \citep{Nishida_1998}. In the last row, $\alpha$ denotes the continuum-level slip magnitude, which is arbitrary, non-zero. See also Figure \ref{mechs} for visualizations of the individual mechanisms.}
\begin{tabular}{p{6cm}cccc}
\hline 
Denotation$^a$ &twinning&invariant&shearing
&shear\\  \addlinespace[-12pt]
{}&type&plane&direction&magnitude\\
\hline
$(001)_{\rm M}$ twins &{compound}& $(001)_{\rm M}$& $[100]_{\rm M}$&0.2385\\ 
$(100)_{\rm M}$ twins &{compound}& $(100)_{\rm M}$& $[001]_{\rm M}$&0.2385\\
$(20\bar{1})_{\rm M}$ twins &{compound}& $(20\bar{1})_{\rm M}$& $[\bar{1}0\bar{2}]_{\rm M}$&0.4250\\ 
{}&{}&{}&{}&{}\\
{$[100](001)_{\rm M}$ dislocation slip}&{-- -- --}&$(001)_{\rm M}$& $[100]_{\rm M}$&$\alpha\in{\mathbb R}$ \\
\hline
\multicolumn{5}{p{\textwidth}}{$^a${\footnotesize In figures, for better visibility we often drop the subscript ${\rm M}$ from the denotations, using e.g. $(100)$~twins instead of $(100)_{\rm M}$~twins, etc.}}
\end{tabular}
\label{tabul}
\end{table}

\begin{figure}[!t]
 \centering
 \includegraphics[width=\textwidth]{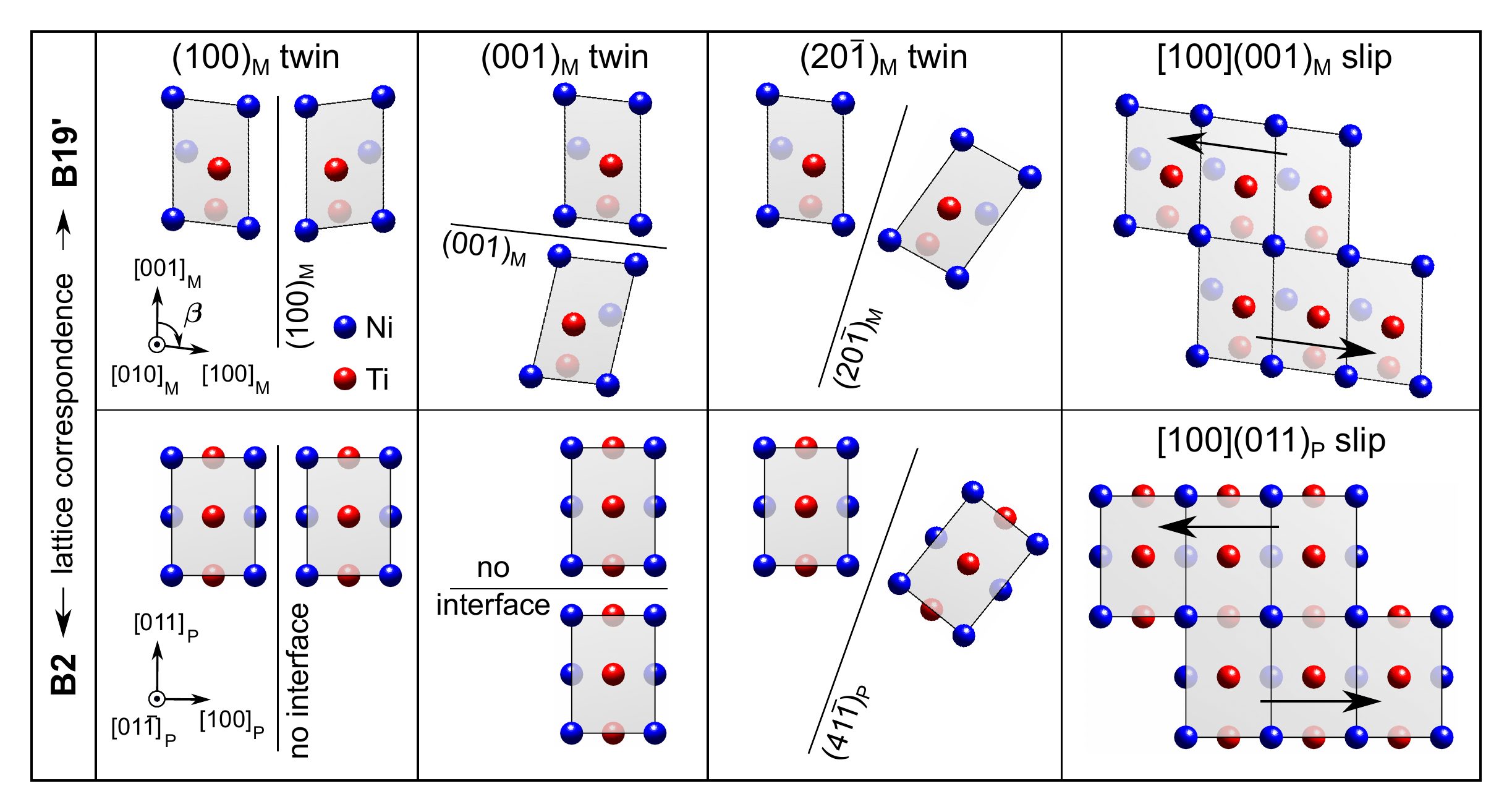}
 \caption{The considered deformation mechanisms acting on one $(010)_{\rm M}$ lattice plane. The upper row shows these mechanism in the B19$^\prime$ martensite lattice, the lower row are the related processes in the parent phase (B2) lattice linked through the lattice correspondence. For the mechanisms within one Eriksen-Pitteri neighborhood (the first two columns), the reference austenite lattice remains unaltered; for the mechanisms straining the lattice outside EPN (the third and the fourth columns), lattice defects and plastic strains are introduced into austenite.}
 \label{mechs}
\end{figure}

From the four deformation mechanisms listed in Table \ref{tabul}, only the $(001)_{\rm M}$ twins and $(100)_{\rm M}$ twins do not affect the corresponding lattice of austenite, i.e., only the strains due to activity of these two mechanisms can be erased by the reverse transition to austenite. This is visualized in Figure \ref{mechs}. The $[100](001)_{\rm M}$ dislocation slip acts as a $[100](011)_{\rm P}$ slip in the corresponding austenitic lattice. The $(20\bar{1})_{\rm M}$ twins correspond to $(41\bar{1})_{\rm P}$ twins in the parent lattice. In agreement with \cite{Gao,pathways}, we use the term 'twinning beyond the Eriksen-Pitteri neighborhood (EPN)' to denote the twinning that affects the corresponding austenite lattice. 

The visualization in Figure \ref{mechs} is created using lattice parameters from \citep{Bhattacharya} for the shapes of the unit cells, and the positions of the atoms inside of the cells follow the atomic coordinates determined experimentally by \cite{Kudoh_1985}. It is worth noting that, unlike for the shape of the unit cell, the inner structure in the B19$^\prime$ cell is not invariant with respect to a 180$^\circ{}$ rotation about the $[010]_{\rm M}$ axis. Consequently, for the visualization we need to choose one of two equivalent orientations of the inner structure; these two orientations represent the same crystal structure, but the location of the unit cell in it differs by a translation vector. For the choice done for Figure \ref{mechs}, there is a full mirror symmetry between the visualizations of the adjacent variants for $(100)_{\rm M}$ twining, and a pseudo-mirror symmetry for $(001)_{\rm M}$ twinning, where the shape of the cell is mirror-reflected, while its inner structure is not. 

Detailed considerations of whether the atomic positions (and the corresponding motifs) are only mirror-reflected, or also translated play an important role for first-principles calculations of the energy of the twins \citep{Ezaz_Acta_2011}, but do not affect the continuum-mechanics construction in this paper, as the shape strain (i.e., the quantity that copies from the atomistic level to continuum through the Cauchy-Born hypothesis \citep{BhatLamin}) in both situations is the same.

\subsection{Summary of experimental observations}

The purpose of this subsection is to summarize two most comprehensively documented case studies on B19$^\prime$ plastic deformation in literature, which are \citep{Zhang} and \citep{Molnarova_2020,Sittner_SMSE_2023_SMST}. We focus on features frequently observed in the experiments, mainly those related to $(20\bar{1})_{\rm M}$ twinning and pattern formation.  Importantly, although these observations were done on very different materials (in terms of chemical composition, grain size, yield strength, etc.), the sequence of the observed microstructures included very similar \emph{initial states} (mixtures of finely $(001)_{\rm M}$-twinned domains in self-accommodated, thermally induced martensite), very similar \emph{intermediate states} (nearly or fully homogeneous fine $(001)_{\rm M}$ lamination when the material was approaching the reversibility limit under tensile loading), and very similar \emph{terminal states} ($(20\bar{1})_{\rm M}$- and $(100)_{\rm M}$-oriented bands, mostly without the $(001)_{\rm M}$-twinned substructure, in the plastically formed material). The similarity in the intermediate state is of a particular importance: in this state, the whole grain or a large area of it has one orientation of the $(010)_{\rm M}$ plane with respect to the loading, and thus, the grain can deform through the two-dimensional deformation mechanisms introduced in Figure \ref{mechs}. The difference can be seen in the reasoning for the appearance of the fine $(001)_{\rm M}$ lamination at the initial and intermediate stages: while  \cite{Zhang} ascribed this lamination to the Ti-excess and Guinier-Preston zones, \cite{Molnarova_2020} have proved that this lamination may originate from the strain-compatibility conditions between neighboring grains. Additional differences were also in some finer details in how the microstructure evolved between the initial, intermediate and terminal stages, which we ascribe mainly to the different grain size, as discussed below. However, these observations are sufficiently similar to justify our ambition to capture them both by the same microstructural model. {{} Let us also remark that in the thermally-induced martensite in NiTi polycrystals, Type II twins are typically observed  as the most prevalent twinning mode \citep{Matsumoto_Acta_1987,Onda_MatTrans_1992}, since these twins can help to satisfy compatibility conditions at the austenite-martensite interface during the transition \citep{Hane_Acta_1999,Bhattacharya}. In nanograined polycrystals, however, the bands holding the Type II (or Type I) twin-like relationships are further internally $(001)_{\rm M}$-compound twinned \citep{Waitz_MST_2008,Molnarova_2020}. This internal twinning further reduces the shape strain of the microstructure, and possibly also leads to some elastic softening of the lattice \citep{Wang_IJP_2014,Grabec_Acta_2021} or results from the rhombohedral R-phase forming during the transition \citep{Zhang_MSEA_2004}. The \emph{intermediate state} is then reached when the Type II twins are removed by reorientation, while the finer-scale $(001)_{\rm M}$ compound twins persist in the grains \citep{Zhang,Chowdhury,Molnarova_2020,Sittner_SMSE_2023_SMST}.}

\medskip
\subsubsection{Observations on melt-spun ribbons}

\begin{figure}[!t]
 \centering
 \includegraphics[width=\textwidth]{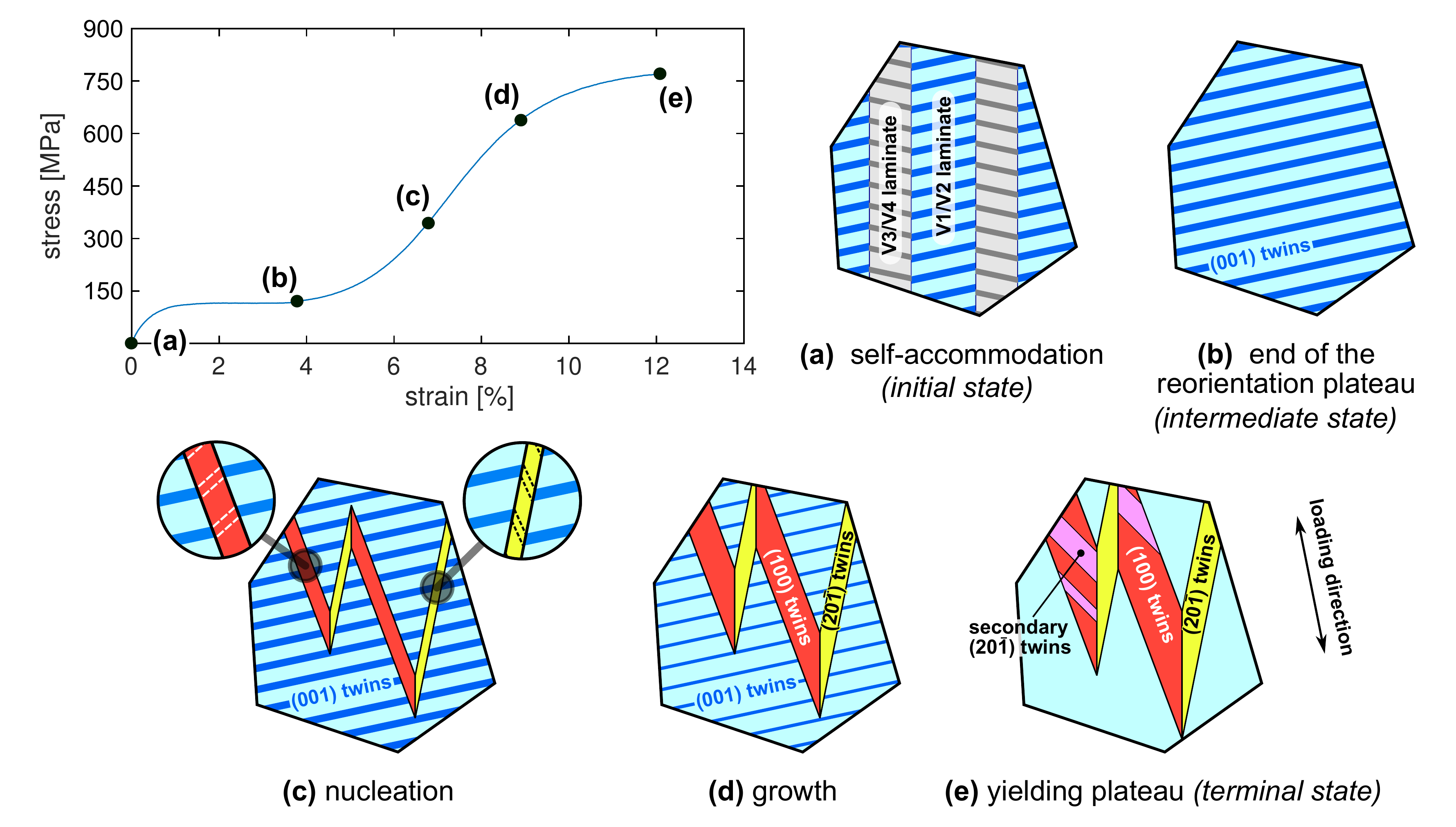}
 \caption{A typical tensile stress-strain curve and the corresponding sequence of microstructures appearing in plastically formed large-grain  NiTi martensite, as observed by electron microscopy \citep{Zhang}: (a) a self-accommodated microstructure originating from the thermally-induced transition; (b) reversibly strained martensite with fine $(001)_{\rm M}$ compound lamination; (c) nucleation of V-shaped microstructures in the laminate with a lamination-like contrast in the newly formed bands; (d) thickening of the new bands and disappearance of the laminate; (e) a final stage with fully formed V-shaped patterns in a single-variant matrix and secondary twinning inside of the bands.  For better visibility, the microstructural features (lengthscales $\sim$ 10-50 nm) are not drawn in scale with respect to the grain size ($\sim$ 5 $\mu$m).  \cite{Zhang} also reported on minor appearance of $(\bar{1}\bar{1}3)$ twins at the very terminal stages, which we do not consider here; we can speculate that these twins appear because of strain accommodation in direction perpendicular to the considered $(010)_{\rm M}$ plane.} \label{evolution}
\end{figure}

\cite{Zhang} reported on microstructure evolution in plastically strained melt-spun ribbons (in form of thin films). They observed a typical sequence of microstructures appearing with an increasing load; this sequence, outlined in Figure \ref{evolution}, was later discussed by  \cite{Chowdhury} and many others. The sequence starts from thermally-induced martensite (Figure \ref{evolution}), consisting of a self-accommodated mixture of variants, often having $(001)_{\rm M}$ compound twins at the finest scale. When this mixture is subjected to uniaxial tension, it first undergoes martensite reorientation at an approximately constant stress level; this stress level defines the so-called reorientation plateau. At the end of the reorientation plateau, the grains are typically full of fine $(001)_{\rm M}$ compound twins (Figure \ref{evolution}(b)). With further loading the  material deforms elastically or by reversible detwinning, until the plastic twins start nucleating in form of V-shaped microstructures.  At the nucleation stage (Figure \ref{evolution}(c)), the $(20\bar{1})_{\rm M}$ and $(100)_{\rm M}$ twin bands include a fine band-like contrast that resembles the herring-bone pattern in thermally induced martensite and provides a continuous connection of the $(001)_{\rm M}$ compound laminate over the newly nucleating bands. This nucleation of the first V-shaped patterns marks the onset of the plastic deformation and the loss of reversibility of the strains. With a further increase of the load, the $(20\bar{1})_{\rm M}$ and $(100)_{\rm M}$ twin bands grow, which is accompanied by a simultaneous disappearance of the fine  $(001)_{\rm M}$ laminate (Figure \ref{evolution}(d)) surrounding the V-shaped microstructures, as well as of the finer contrast inside of them. Finally, the material reaches the plastic-forming plateau (termed hereafter the yielding plateau), where the plastic deformation proceeds via growth of the  $(20\bar{1})_{\rm M}$ and $(100)_{\rm M}$ twin bands and appearance of secondary twins inside them (Figure \ref{evolution}(e)).

The stages (a-e) in Figure \ref{evolution} may appear at different stress levels and for different strains, depending on temperature and on the particular microstructure and chemical composition of the alloy. For Ti-rich Ni-Ti  thin ribbons (Ni$_{48.9}$Ti$_{51.5}$, grain size $\sim$ 5 $\mu$m) reported by \cite{Zhang}, the lower plateau appeared at approximately 150 MPa and the stage (b) was reached for axial strain $\varepsilon\approx$ 4\%. The upper plateau (stage (e)) was at approximately 700 MPa and $\varepsilon\approx$ 12\%; the stages (c) and (d) correspond to strains of $\varepsilon\approx$ 7\% and $\varepsilon\approx$ 9\%, respectively.  

The coexistence of $(20\bar{1})_{\rm M}$ and  $(10{0})_{\rm M}$ twins in form of the V-shaped microstructure  was reported also for hot-forged and drawn polycrystalline NiTi rods \citep{Ii}, for cold-rolled and annealed NiTi sheets \citep{Nishida_1998} or for a single crystal \citep{Ezaz_MSEA_2020}; patterns clearly resembling the $(20\bar{1})_{\rm M}$/$(10{0})_{\rm M}$ wedges were seen also in plastically formed residual bands inside of austenite grains in hot-rolled NiTi sheet with a more than 20~$\mu$m average grain size \citep{Polatidis_2020}. Several other works report on $(41\bar{1})_{\rm P}$ twin patterns appearing in austenite after a plastically formed B19$^\prime$ martensite had been transformed to the parent phase due to unloading and/or heating \citep{Nishida_2006,Sittn1}. These patterns are very similar to those of in Figure \ref{evolution}(c), because they are inherited from the martensite phase due to the crystallographic equivalence between the $(20\bar{1})_{\rm M}$ and $(41\bar{1})_{\rm P}$ twins, outlined in the third column of Figure \ref{mechs}.

\subsubsection{Observations on cold-worked $(111)_{\rm A}$-textured wires}

A detailed transmission electron microscopy (TEM) analysis of microstructure evolution under plastic forming has been recently reported by  \cite{Molnarova_2020} for strongly textured NiTi wires (Ni$_{49.5}$Ti$_{50.5}$ with a grain size of $\sim$ 0.2 $\mu$m). The TEM analysis has been complemented with high-resolution TEM (HRTEM) in \citep{Sittner_HRTEM_SSR} (see also the Supplementary material in \citep{Molnarova_2020}), supported by texture evolution observations \citep{Bian_AppMatToday_2022}, and comprehensively summarized recently in \citep{Sittner_SMSE_2023_SMST}.

The pattern formation was observed on the $(010)_{\rm M}$ crystallographic plane in each grain, and, except of for the self-accommodated thermally-induced martensite at the very beginning of the experiment, it utilized only the straining mechanisms acting on this plane and summarized in Figure \ref{mechs}. For the tests performed at   20 $^\circ$C, the reorientation plateau was at approximately 150 MPa again, but the upper plateau was close to 1100 MPa. The axial strains corresponding to (b) and (e) were 7 \% and 15 \%, respectively. When the same experiments were performed at 100 $^\circ$C, where the martensite is stress-induced, the plateaus appeared at 400 MPa and 825 MPa, respectively, but exactly the same V-shaped patterns were observed in the material deformed to 15 \%. This proves that the mechanism associated with the formation of these patterns is general, and probably is one of the main mechanisms of plastic forming in NiTi.  Unlike for the melt-spun ribbons in \citep{Zhang}, no nucleation of $(20\bar{1})$-oriented bands below the yielding plateau has been observed for the cold-worked wires, that is, the stage (c) in Figure \ref{evolution} was absent. This can be rationalized by the grain size, as smaller grains pose higher barriers for formation of the microstructure \citep{Waitz_JMPS_2007,Waitz_MST_2008,Kabla_Acta_2014}, and thus, it appears at much higher driving forces. For this reason, while in the melt-spun ribbons the onset of irreversible strains and appearance V-shaped patterns were spread over the the whole stress-strain curve above the point (c), for the cold-worked wires they became active at the yielding plateau only.

Furthermore, the analysis reported in \citep{Molnarova_2020} revealed some dependence of the pattern formation on the grain orientation with respect to the loading axis. Nearly all grains were finely $(001)_{\rm M}$-twinned at the end of the reorientation plateau, which was rationalized in \citep{Molnarova_2020} by the compatibility conditions at the grain boundaries for the given $(111)_{\rm P}$ wire texture. This $(001)_{\rm M}$ lamination in all grains with one major variant in each grain has been confirmed also by texture measurements \citep{Bian_AppMatToday_2022}. Nevertheless, further evolution of the microstructure was different for the grains oriented exactly along the texture, and those specifically inclined from the major orientation. In particular, for ${\mathbf p}$ denoting the projection of the loading axis onto the observed $(010)_{\rm M}$ plane, the following dichotomy was observed:

\begin{enumerate}
  \item{For the grains where ${\mathbf p}$ lied approximately perpendicular to the $(001)_{\rm M}$-compound twinning planes, the plastic-forming sequence was very similar to the one observed in melt-spun ribbons, with a massive appearance of V-shaped microstructures, and their co-existence with gradually disappearing $(001)_{\rm M}$ laminates. For this direction of the vector ${\mathbf p}$, the variants forming the  $(001)_{\rm M}$-laminate were energetically equivalent with respect to the loading; for this reason, some $(001)_{\rm M}$-twinned regions outside the $(20\bar{1})_{\rm M}$ and $(100)_{\rm M}$ twin bands persisted in the microstructure even at the yielding plateau. At the same time, some deformation bands observed in the grains were not twin bands, but bands of rotated lattice (so-called \emph{kink bands}, Figure 7 in \citep{Molnarova_2020}) as usual in plastically formed materials with highly anisotropic plastic slip 
\citep{Hess_TAIME_1949,Inamura}. The presence of kink bands confirmed the massive activity of the $[100](001)_{\rm M}$ dislocation slip in these grains.}
  
  \item For the grains where ${\mathbf p}$ lied approximately along the $[10\bar{1}]_{\rm M}$ direction of the major variant in the $(001)_{\rm M}$-compound laminate, only few V-shaped microstructures were observed, while the dominant objects were the $(20\bar{1})_{\rm M}$ twin bands. In contrast, no isolated $(100)_{\rm M}$ twin bands were observed. In these grains, one of the variants forming the $(001)_{\rm M}$ laminate is strongly energetically preferred by the loading; this laminate was completely absent in microstructures representing the material strained beyond the start of the yielding plateau. In other words, the main mechanism of the plastic forming along the yielding plateau in these grains was the growth of $(20\bar{1})_{\rm M}$ plastic twin bands in an otherwise fully detwinned oriented martensite.
 
\end{enumerate}

For the former case (${\mathbf p}$ perpendicular to $(001)_{\rm M}$), the microstructures were becoming more and more complex with a further progress of plastic forming \citep{Molnarova_2020, Sittner_HRTEM_SSR, Sittner_SMSE_2023_SMST}, until the grain became finally fully filled with various deformation bands. The planar interfaces between the bands were identified as either providing the $(100)$ or $(20\bar{1})$ twin relations between the neighboring lattices, or being just tilt interfaces between two differently rotated lattices belonging to the same variant of martensite.

\bigskip
Another important observation of \cite{Molnarova_2020, Sittner_HRTEM_SSR, Sittner_SMSE_2023_SMST}
were the orientation relationships between the lattices forming the  $(20\bar{1})_{\rm M}$ twins, and the resulting geometry of the $(20\bar{1})_{\rm M}$-twin bands. For compound-type twins lying along a crystallographic plane and representing a shear along a lattice direction, one would expect a quite strictly defined orientation relationship and small deviations from the exact twinning plane only. Instead, a quite strong scatter from the theoretically predicted orientations was reported (see \citep{Sittner_HRTEM_SSR}, we discuss this in more detail in Section 4), also there was a considerable disorder seen in the crystal structure surrounding the $(20\bar{1})_{\rm M}$-twin interfaces seen in the HRTEM images \citep{Ii,Sittner_HRTEM_SSR}. The misorientations and local disorders are then also inherited in austenite during the reverse transition. As mentioned in \citep{Molnarova_2020}, the V-shaped patterns formed by $\{41\bar{1}\}$ twins in austenite also deviate from being attached to exact lattice planes and there is some scatter in the tilt angles between the lattices; The diffuse character of $\{41\bar{1}\}$ twins is seen in \citep{Ii}. All these observations indicate that the character of the $(20\bar{1})_{\rm M}$ twins in martensite might be somehow different from other compound twins commonly observed in shape memory alloys, revealing, again, the need of a more detailed analysis of their origin and their contribution to plastic forming of NiTi.

\subsection{{}{Motivation for the kwinking model}}

The theoretical model capturing the above summarized experimental observations is in detail formulated in Section 3. Here, we explain the motivation for main features of the model, using particular examples from the experiment. This subsection gives a heuristic illustration of how essential is the role of the dislocation slip and how the slip can be coupled with the twinning; in Section 3, we discuss how this importance of the slip and the coupling can be incorporated in a unified, general model.

\subsubsection{Nucleation and growth of the deformation bands in a $(001)_{\rm M}$-compound laminate}

\begin{sidewaysfigure}[p]
 \includegraphics[width=\textwidth]{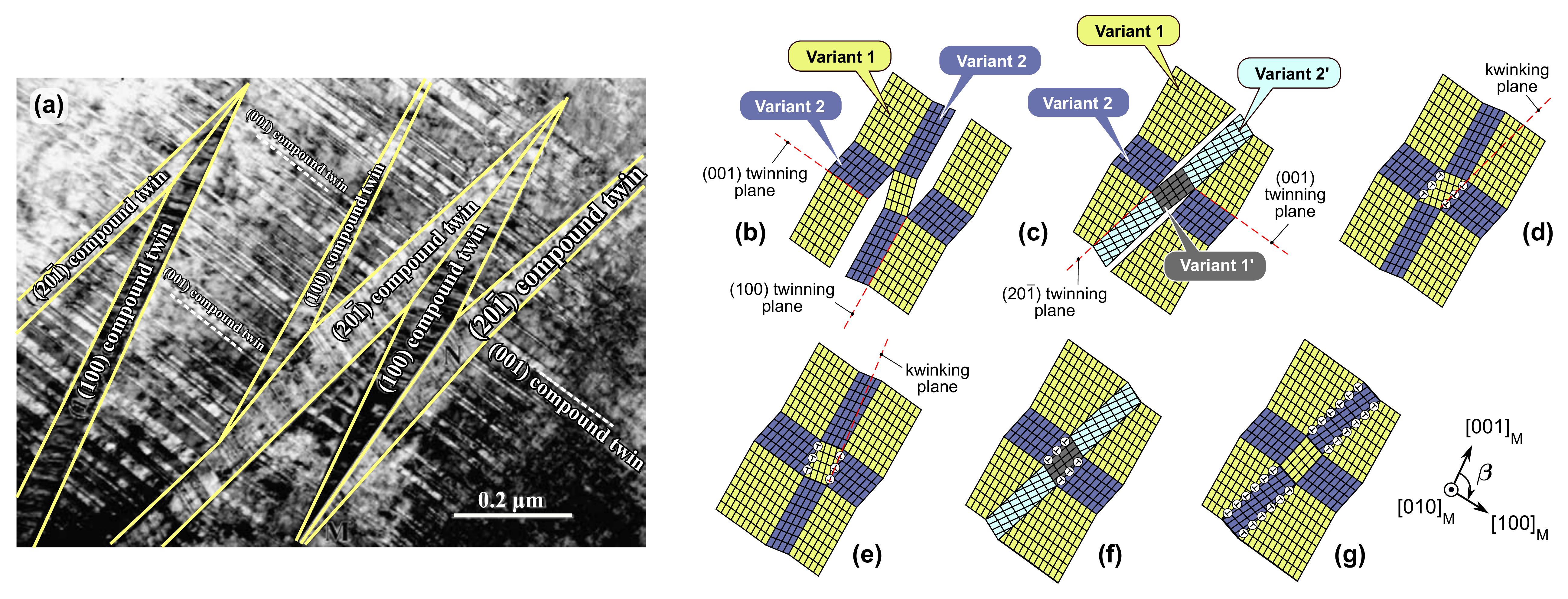}
 \caption{{}(a) an experimental micrograph of $(100)_{\rm M}$ and $(20\bar{1})_{\rm M}$ twin bands crossing with the initial $(001)_{\rm M}$ twin laminate, micrograph modified after \cite{Zhang} with permission from Elsevier; (b,c) visualization of the incompatibility resulting from the crossing; (d,e,f,g) visualizations of the compensation of the incompatibility through a $[100](001)_{\rm M}$ dislocation slip.  The structure in (d) has been created from (b) by dislocation slip in the intersection area only, assuming that all other regions and their boundaries remain unaltered; in (e), additional energy relaxation is enabled through allowing martensite reoreintation in these other regions. In (b-f), lattice parameters of B19$\prime$ introduced in Section 2.1 were used for the visualizations; the kwinking plane orientations in (d-g) are just illustrative and approximate, their exact parameters are calculated in section 4.1.}
 \label{pasypopisy}
\end{sidewaysfigure}

First we focus on the nucleation stage of the V-shaped patterns in melt-spun ribbons (Figure \ref{evolution}(c)), where the $(20\bar{1})_{\rm M}$ and $(100)_{\rm M}$ twin bands are formed inside of a $(001)_{\rm M}$-twinned lattice. As clearly documented by  \cite{Zhang}, both $(20\bar{1})_{\rm M}$ and $(100)_{\rm M}$ bands cross the fine $(001)_{\rm M}$-compound twins existing at the end of the reorientation plateau, i.e., the newly formed bands are also internally twinned and there is a continuous, one-to-one connection between the twins inside the bands and the $(001)_{\rm M}$-compound twins in the matrix (as seen in several places in Figure \ref{pasypopisy}(a)). From their crystallographic orientations, and also because there is no other reversible twinning system available in the given plane, it is clear that the fine twins inside the bands are, again, $(001)_{\rm M}$-compound twins. Hence, the first requirement we put on the model (in the heuristic approach) is that the model must be able to capture the compatible connections of $(001)_{\rm M}$-compound twins over the $(20\bar{1})_{\rm M}$ and $(100)_{\rm M}$ twin bands that nucleate at the very first stages of plastic forming.

As shown by the mathematical theory of martensite, the so-called \emph{crossing-twins microstructure} is energetically admissible only if the twinning components of the involved twins satisfy a quite strict set of algebraic conditions \citep{Bhattacharya,JMPS}. The set is treated in Section 4.1 by an explicit calculation. However, to show that this set is not  satisfied (not even in a approximate sense) for a combination of $(001)_{\rm M}$ and $(100)_{\rm M}$ compound twins from one EPN, and also if the band undergoes exact $(20\bar{1})_{\rm M}$ twinning outside the given EPN, it is fully sufficient to use visualizations, as done in Figures \ref{pasypopisy}(b) and (c). Figure \ref{pasypopisy}(b) shows a matrix (denoted as Variant 1) which includes two  bands of Variant 2 that are expected to cross each other. One band holds a $(001)_{\rm M}$ twin relation with the matrix and the second one the $(100)_{\rm M}$ twin relation. It is seen that the area where the bands intersect cannot hold twin relations to the lattices inside both bands and a geometric misfit (i.e., an incompatibility) appears. A similar situation appears for a $(20\bar{1})_{\rm M}$ twin band, as shown in Figure \ref{pasypopisy}(c). Here, the major variant of the band is some Variant 2$^\prime$, which belongs to another EPN and holds a $(20\bar{1})_{\rm M}$ twin relation to the matrix; the $(001)_{\rm M}$ twin relation in this EPN transforms Variant 2$^\prime$ into Variant 1$^\prime$. Again, the Variant 1$^\prime$ in the area where the bands intersect is not compatibly connected to the $(001)_{\rm M}$ band in the matrix, and there is an incompatibility. 

For close-to-compatible microstructures, where the misfit angle is $\lesssim$1$^\circ$ \citep{Balandraud_JMPS_2007,Heczko_Acta_2013}, the compatibility can be achieved by elastic strains. Here, however, the misfit angles are $\approx$27$^\circ$ for the situation sketched in Figure \ref{pasypopisy}(b) and $\approx$10$^\circ$ for the situation sketched in Figure \ref{pasypopisy}(c). Hence, plastic slip is required to compensate the misfit. The B19$^\prime$ martensite has only one slip system available, which is the $[100](001)_{\rm M}$ slip; however, this one slip system is enough for the Variant 1 inside of the area where the bands intersect to reach such strain that the incompatibility is significantly reduced (although not fully removed, as discussed below). This situation is sketched in Figure \ref{pasypopisy}(d): the new microstructure consists of $(001)_{\rm M}$ and $(100)_{\rm M}$ twins, plus there are dislocation cores appearing between Variant 1 and Variant 2 that result from the plastic slip.  

The newly formed planar interface between Variant 2 and plastically slipped Variant 1 requires some additional comments. It is a twin interface with an additional rotation resulting from the presence of the dislocation cores. Hence, this interface combines the character of tilt interfaces that typically encapsulate \emph{kink bands} in layered media or media with highly anisotropic dislocation slip (such as long period stacking order magnesium alloys) with the twin interfaces that encapsulate \emph{twin bands}, as usual in shape memory alloys. We will later show that such combined interfaces and such combined \emph{kink-twin} bands may play a crucial role in the discussed plastically formed patterns in NiTi. Because of the lack of any established terminology, we call these interfaces \emph{kwink interfaces} (or simply \emph{kwinks}, which comes from combining the words \emph{twin} and \emph{kink}), and we call the resulting bands the \emph{kwink bands}. 

 The situation in Figure \ref{pasypopisy}(d) is idealized, assuming that while the material in the intersection area becomes plastically sheared, all other regions remain intact and are just translated to stick to each other in the resulting microstructure. This leads to just a partial relaxation of the incompatibility; the material still needs to be elastically strained to reach the compatibility conditions over the kwink interfaces. The reason is that the extrapolations of the $(001)_{\rm M}$ slip planes in the intersection area do not exactly meet the $(001)_{\rm M}$ planes in the rest of the $(001)_{\rm M}$-twin band. However, the microstructure now has additional degrees of freedom, through which the full compatibility can be reached: the $(100)_{\rm M}$ twin band in the lower part of the domain can move to the left or to the right by martensite reorientation, and the kwink interfaces can tilt accordingly, ensuring the continuity of the band. As shown in Section 4.1, the full compatibility is achieved if the lattices forming a kwink hold a twin-like relation which is quite close to a $(20\bar{1})_{\rm M}$-twin relation, and also the orientation of the resulting kwink band is quite close to the $(20\bar{1})_{\rm M}$ plane. This is approximately drawn in Figure \ref{pasypopisy}(e), where the kwinking planes are exactly $(20\bar{1})_{\rm M}$ with respect to the lattice of Variant 2. In other words, the slip inside the $(001)_{\rm M}$ can enable a fully compatible, i.e., energetically cheap, penetration of this band by a $(100)_{\rm M}$ twin band, and then, as a result, new interfaces running along planes close to $(20\bar{1})_{\rm m}$ form, possessing mirror symmetry between the neighboring lattices. Hence, we observe that the $(20\bar{1})_{\rm M}$-twin relation can be (approximately) achieved without any $(20\bar{1})_{\rm M}$-twinning mechanism overcoming the barrier between different EPNs, but, instead, by combining two very simple straining mechanisms, which are the $(100)_{\rm M}$ compound twinning and the $[100](001)_{\rm M}$ slip. The reason is that the coordinated $[100](001)_{\rm M}$ slip, in fact, acts as a mechanism that translates the given material point between different EPNs.

For the case in Figure \ref{pasypopisy}(f), a similar compensation of the incompatibility is possible by activation of the plastic slip in Variant 1$^\prime$. Again, a kwink interface forms between Variant 1$^\prime$ and Variant 2, and the orientation relationship between these two variants over the kwink interface is quite close to the $(100)_{\rm M}$ compound twin relation. Nevertheless, in the light of the observation done for Figures \ref{pasypopisy}(b) and \ref{pasypopisy}(d), we can alternatively assume that the discussed morphology may arise without the $(20\bar{1})_{\rm M}$ twinning, i.e. by combining reversible twinning with plastic slip. Notice that the nucleation of the $(100)_{\rm M}$ bands and the $(20\bar{1})_{\rm M}$ bands appears at the same stress level, which means that if the construction in Figures \ref{pasypopisy}(b) and \ref{pasypopisy}(d) is correct, the slip is already massively active at this level, and there is no reason why it could not be creating the $(20\bar{1})_{\rm M}$-oriented bands themselves. Then, as sketched in Figure \ref{pasypopisy}(g), there is a kwink band with an approximately $(20\bar{1})_{\rm M}$ orientation, that can compatibly cross with a $(001)_{\rm M}$ twin band. Again, the microstructure is composed just of Variant 1 and Variant 2, plus there is a $[100](001)_{\rm M}$ plastic slip, and thus, the situations outlined in Figures \ref{pasypopisy}(d) and (g) are utilizing the same mechanism to enable a compatible crossing of a $(001)_{\rm M}$ twin band with a $(100)_{\rm M}$ twin band and a $(20\bar{1})_{\rm M}$ kwink band, respectively. As a result, the whole microstructure observed in Figure \ref{pasypopisy} can be constructed by combining the slip-enabled compatible crossings from Figures \ref{pasypopisy}(d) and (g); in other words, introducing the kwinking enables explaining this microstructure in its whole richness just by combining three simple deformation mechanism: $(100)_{\rm M}$ compound twinning, $(001)_{\rm M}$ compound twinning, and $[100](001)_{\rm M}$ dislocation slip.

Because of formulating the model at the continuum length-scale (with large strains and large rotations considered, as usual for the mathematical theory of martensitic microstrutures \citep{JMB1,JMB2,Bhattacharya}), we do not discuss here the mechanism the penetration of $(100)_{\rm M}$ twin bands by $(001)_{\rm M}$ in the topological sense, that is, in terms of twinning dislocations and twinning disconnections that interact at the interfaces. That would definitely be needed for any more quantitative modelling, considering the energy barriers for the penetration as well as the driving forces on the individual line defects. The penetration models at this level are well developed in literature, with the main ideas owing to  \cite{Mullner_MSEA_1997,Mullner_ZMk_2006}. In these topological models, the incompatibility visualized in Figure \ref{pasypopisy}(b,c) would be represented by a pair of disclination dipoles in an otherwise connected material, and the dislocation walls appearing at the kwink interface would be understood as arrays of defects, the stress fields of which compensate the dipoles (cf. \citep{Mullner_Acta_2000,Mullner_Acta_2010,Mullner_JALCOM_2013}). Such a description would be, in many respects, fully equivalent to our all-continuum approach, possibly elucidating more details on how the line defects responsible for the growth of the $(001)_{\rm M}$ and $(100)_{\rm M}$ twins (i.e., the twinning dislocations) couple with the slip dislocations into new line defects substantiating the nucleation and growth of the kwink interfaces. At the same time, with a topological model we would loose the advantage of representing all active deformation mechanisms by simple two-dimensional deformation gradients, as done in Section 3. Nevertheless, the concept (and terminology) of disclinations and their compensation is applicable also at the continuum level, and we will utilize it Sections 2.3.4 and 4.3, when discussing the formation of the V-shaped patterns.

\begin{figure}
 \centering
 \includegraphics[width=0.95\textwidth]{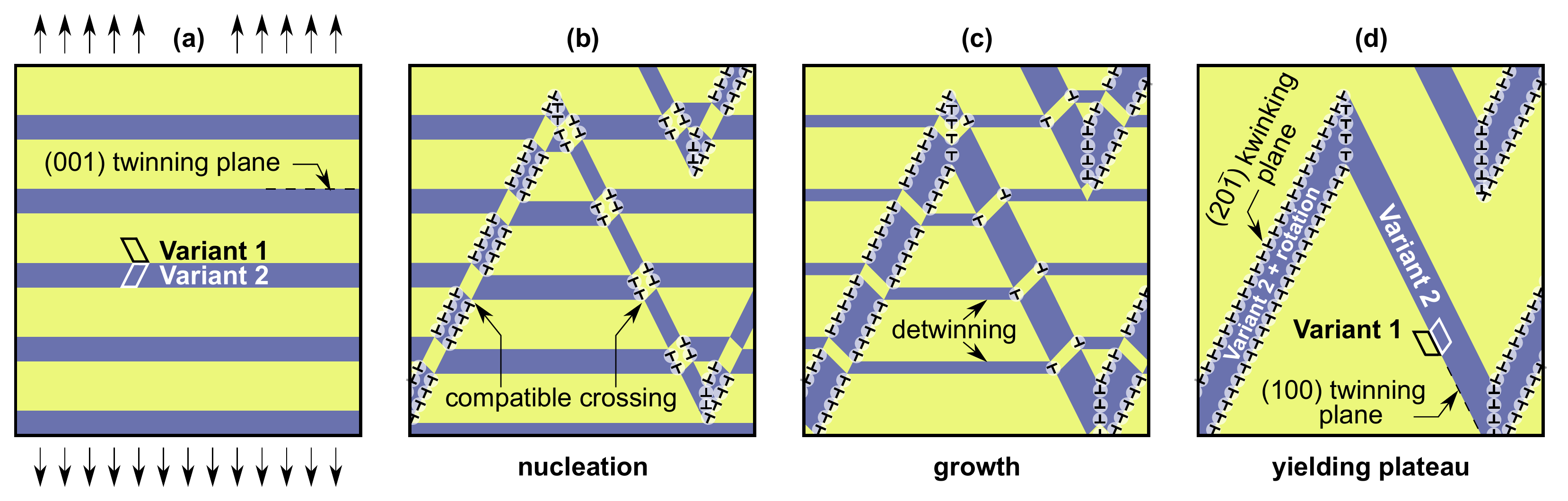}
 \caption{{}A simplified evolution of the microstructure from a $(001)_{\rm M}$ laminate (a) into the V-shaped patterns (d), as suggested for the demonstration of the main features of the developed model. The compatible crossings in (b) are enabled by the $[100](001)_{\rm M}$ slip inside of the bands; after the detwinning (c), the microstructure is composed just of $(001)_{\rm M}$ twin bands and $(20\bar{1})_{\rm M}$ kwink bands, the latter being decorated by dislocation cores. For simplicity and better visual clarity, this figure is using a schematic geometry, not utilizing real shape strains or real orientation relations in the lattice.} \label{zigzage}
\end{figure}

\bigskip
 The kwinks, as introduced above, allow for compatible crossing between the $(001)_{\rm M}$ and $(100)_{\rm M}$ twins. This enables us to interpret the transition from the $(001)_{\rm M}$ laminate into the pattern consisting of $(100)_{\rm M}$ twins and $(20\bar{1})_{\rm M}$ kwinks, as described in Section 2.2. This interpretation is outlined in Figure \ref{zigzage}. We consider first a $(001)_{\rm M}$ compound twin laminate, consisting of two variants, Variant 1 being the major one. The laminate is inside a  material loaded in tension beyond the reorientation plateau. For simplicity, we assume that the loading direction is perpendicular to the  $(001)_{\rm M}$ planes; this means that neither the $(001)_{\rm M}$ compound twins, nor the $[100](001)_{\rm M}$ plastic slip can relax the external loads, as these mechanisms have both zero Schmid factors. In $(111)_{\rm M}$-textured NiTi wires, the $(001)_{\rm M}$ compound laminate at the end of the reorientation plateau is stabilized due to compatibility conditions between grains \citep{Molnarova_2020}, and the loading direction points rather along the $[101]_{\rm M}$ lattice vector. That situation is too complex to be captured by our model, but even in such a case, there is a large component of the loading that is perpendicular to the $(001)_{\rm M}$ planes, and the $(001)_{\rm M}$ compound twins and the $[100](001)_{\rm M}$ plastic slip cannot fully relax the external loads, so the sequence outlined in Figure \ref{zigzage} may still give a quite good approximation of what happens in individual grains. Notice that this sequence explains only how the microstructure in Figure \ref{zigzage}(a) transforms into the microstructure in Figure \ref{zigzage}(d) under the external load, not how the V-shaped patterns accommodate tensile strains in the loading direction, or how does the material tackle stress singularities appearing at the tips of the patterns (i.e. in the regions where the $(100)_{\rm M}$ twins and $(20\bar{1})_{\rm M}$ kwinks meet). This will be discussed separately, first heuristically in subsection 2.3.4, and then by an explicit calculation in section 4. 

\bigskip
For the material in Figure \ref{zigzage}(a) loaded in the vertical direction, the only available mechanism within the given EPN that has a non-zero Schmid factor is the $(100)_{\rm M}$ twinning. And thus, the material needs to utilize this mechanism to relax the stress.  However, the $(100)_{\rm M}$ twin bands need to intersect with the $(001)_{\rm M}$ compound laminate, which is not possible due to compatibility reasons, as discussed above. If we admit that there is a $[100](001)_{\rm M}$ slip active in the microstructure, and that it enables formation of kwink interfaces, there are two types of deformation bands that utilize the $(100)_{\rm M}$ twinning: one holds the $(100)_{\rm M}$-twin relation with the major variant (and becomes a kwink band with respect to the minor variant), and the second one holds the $(100)_{\rm M}$-twin relation with the minor variant (and becomes a kwink band with respect to the major variant). Because they are created by the same mechanism, we can assume that they nucleate simultaneously in the microstructure (Figure \ref{zigzage}(b)); in fact, as these two bands provide shearing along two different planes, inclined symmetrically with respect to the loading direction, their combination is needed to provide tensile strain along this direction.  Bands of both types cross compatibly with the $(001)_{\rm M}$ laminates, which means they do not cost any extra elastic energy, and also their faces are partially composed of $(100)_{\rm M}$ twinning planes, so they are cheaper in terms of the surface energy than any kink or kwink bands with other orientations (see Section 3.3). To nucleate the bands, the external stress must be large enough to trigger a massive, coordinated plastic slip that overcomes the strong plastic anisotropy of the B19$^\prime$ lattice; that is why the yielding starts at much higher stress levels compared with the reorientation plateau.

Once the bands are nucleated, the $(001)_{\rm M}$ planes inside of them are tilted with respect to the loading axis, which leads to non-zero Schmid factors for the  $(001)_{\rm M}$ compound twinning, and creates a driving force for $(001)_{\rm M}$-detwinning, as well as for the $[100](001)_{\rm M}$ dislocation slip. The detwinning inside the bands must be accompanied by detwinning outside bands, because of the compatible crossing between the laminates. Simultaneously, the bands grow (Figure \ref{zigzage}(c)), as both the detwinning inside the bands and thickening of the bands themselves help to accommodate the tensile strains. Finally, the $(001)_{\rm M}$ compound laminate completely disappears (Figure \ref{zigzage}(d)), and further plastic straining would lead to formation of additional V-shaped microstructures, their growth and collisions, and secondary twinning inside of the bands. None of these mechanisms is affected by the compatible crossing between the deformation bands and the $(001)_{\rm M}$ twins anymore; the main morphology of the pattern is, nevertheless, inherited from the nucleation stage, where this feature is essential.

\subsubsection{General comments on the concept of kwinks}

\begin{figure}
 \centering
 \includegraphics[width=\textwidth]{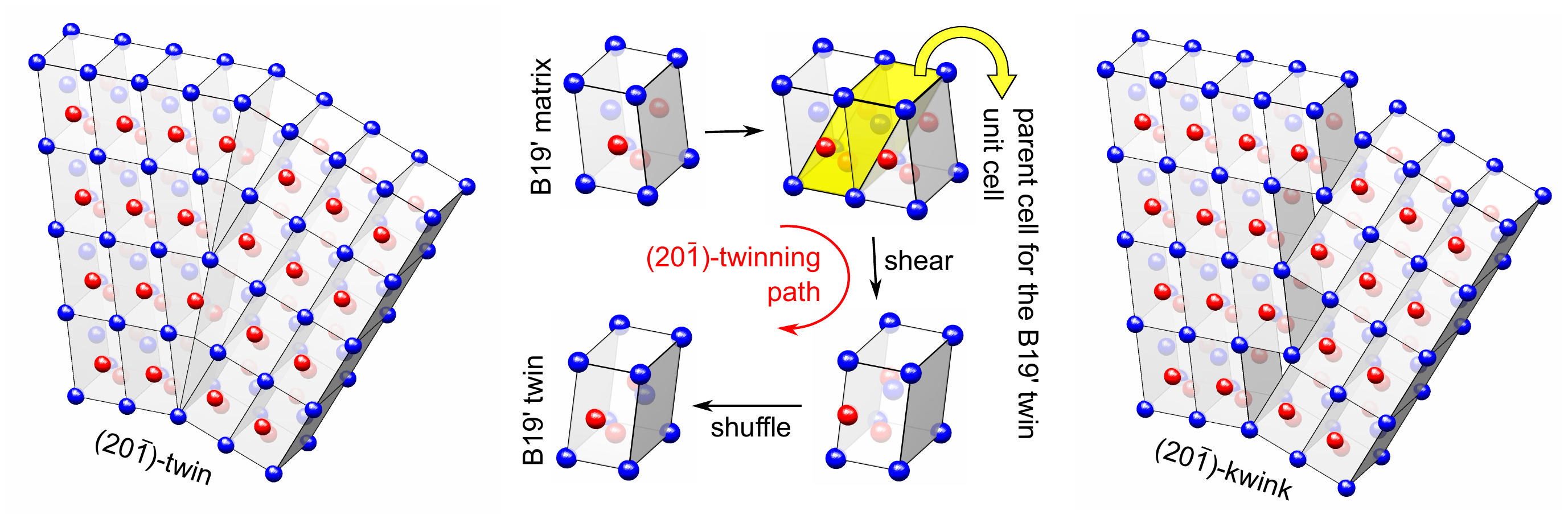}
 \caption{Atomistic-scale visualization of the difference between a $(20\bar{1})_{\rm M}$ twin and a $(20\bar{1})_{\rm M}$ kwink. Both deformation mechanisms result in the same shape strains and the same twin-relationship between the neighboring lattices. However, while for the twin this relationship is achieved through a complex twinning path (shown schematically in the middle, see \citep{Gao,Ezaz_MSEA_2020} for more details), in the kwink the relationship results from an array of dislocation cores arranged along an originally $(100)_{\rm M}$ twin interface.}
 \label{kulicky}
\end{figure}

The advantage of considering the $(20\bar{1})_{\rm M}$ interfaces as kwinks instead of twins for the modelling purposes is quite obvious (and it will be more detailed in Section 3): while the $(100)_{\rm M}$ twinning can be captured by the mathematical theory of martensitic microstructures, and the unidirectional plastic slip by conventional crystal plasticity tools, which are both well-established modelling approaches, the twinning outside the EPN would lead to several open theoretical questions, such as the loss of the reference configuration (cf. \citep{ZanPRL}). Also, the structure and height of the energy barrier between different EPNs is not sufficiently understood yet. Moreover, there also are several indications that the $(20\bar{1})_{\rm M}$ interfaces indeed arise rather by kwinking than by irreversible twinning. At the atomistic scale, the difference between these two approaches is drawn in Figure \ref{kulicky}. For the plastic $(20\bar{1})_{\rm M}$ twinning, the unit cells need to undergo extensive shuffling of the atoms in addition to the shear deformation, where the individual shuffling atoms are expected to move over distances comparable to the lattice spacing (see also \citep{DTW} for a similar twinning path for the $(41\bar{1})$ deformation twinning in B2 austenite); the distances the atoms need to move are by an order of magnitude larger than the shuffling amplitudes commonly accompanying plastic twinning e.g. in hexagonal materials. As shown by \cite{Ezaz_MSEA_2020}, such a large-amplitude shuffling might be energetically very demanding, which makes the $(20\bar{1})_{\rm M}$ twinning energetically disadvantageous compared with the reversible $(100)_{\rm M}$ and $(001)_{\rm M}$ twinning. In contrast, the $[100](001)_{\rm M}$ plastic slip is geometrically simple and energetically quite cheap, and so is the $(100)_{\rm M}$ twinning, so their combination as of two coupled processes might be energetically favorable over the $(20\bar{1})_{\rm M}$ twinning at the atomistic level. 

On the other hand, as discussed in \citep{Gao,pathways}, the barrier between the EPNs might be not as high as predicted in \citep{Ezaz_MSEA_2020}, because the $(20\bar{1})_{\rm M}$-twinning path goes over the orthorhombic structure B33, which is frequently identified as the ground state structure of NiTi by ab-initio calculations \citep{Wagner,Huang_NMAT_2003}. Further atomistic-scale considerations also reveal that the two discussed mechanisms, $(20\bar{1})_{\rm M}$ twinning and kwinking, are essentially very similar at this scale, and might be even seen as just two different descriptions of the same phenomenon: for example, the (yellow) parent cell for the B19$^\prime$ twin unit cell sketched in the middle part of Figure \ref{kulicky} can be turned into the twin unit cell (up to the shuffle) not only by shearing along the $(20\bar{1})_{\rm M}$ plane, but equivalently by slipping along the $(001)_{\rm M}$ plane. And, again, such a slipping passes through a structure which is geometrically very close to B33. {{} At the same time, the kwinking mechanism at the atomistic scale poses several intriguing questions, such as how do the dislocation cores enter and leave the kwinks, or what barriers the kwinks need to overcome during their motion. Resolving these questions would require a detailed atomistic-topological analysis of the kwinks, as done for example for Type II twins in B19$^\prime$ NiTi by \cite{Mohammed_Acta_2020,Mohammed_Acta_2020b,Mohammed_Acta_2021}.}

For these reasons, we restrict further distinguishing between kwinking and twinning only to the continuum scale, where we can understand kwinking as a result of simultaneously appearing martensite reorientation and a massive coordinated $[100](001)_{\rm M}$ plastic slip, and twinning as a separate mechanism, a pure shear strain localized to the twinning plane and oriented along it that can exist independently of the plastic slip in the surrounding material. In this sense, when formulating the model in Section 3, we define \emph{kwink interfaces} as interfaces between different variants of martensite that, at the same time, represent a jump of plastic slip amplitude, and we stick to this definition throughout the rest of the paper. 

\bigskip
At the continuum lengthscale, the heuristic insight into the nature of the observed microstructures comes  from the pattern formation. The V-shaped patterns (Figure \ref{pasypopisy}(a)) clearly resemble the patterns forming in layered media, in particular those observed in a compression-loaded pile of papers by \cite{Wadee}, but also the typical plastic kinks that appear in magnesium alloys \citep{Inamura,Lei}. On the other hand, such patterns are not typically observed in other shape memory alloys or in NiTi not strained beyond the recoverability limit. In these materials, regular 1st or higher order lamination is typically observed, arising from energy minimization \citep{Bhattacharya} and from the nucleation and growth mechanism of the martensite phase \citep{AdvFunMater}. This suggests that the underlying mechanism is similar, i.e., the observed patterns in NiTi martensite arise due to the plastic slip as well. We will discuss the activity of the plastic slip during the formation of the V-shaped patterns in more details in Section 4.2.

 The experimental observations that the orientation relationship over the $(20\bar{1})_{\rm M}$-oriented interfaces is not exactly a $(20\bar{1})_{\rm M}$-twin relation (as discussed in Section 2.1, see also Section 4.2 and \citep{Sittner_HRTEM_SSR} for quantitative data), may be understood as another supporting argument for the existence of the kwinks. In Section 3.3, we will suggest that the surface energy of a kwink may be reduced by increasing  locally the number of coherent sites at the atomistic scale, which creates a driving force trying to align  segments of the kwinks exactly along the $(20\bar{1})_{\rm M}$ planes. Nevertheless, the compatibility conditions, such as those arising from the compatible crossing with the $(001)_{\rm M}$ twins, are probably the leading energy terms here, and thus, at larger than atomic lengthscales the misorientation is observed. Some misorientations may arise also from the fact that the creation of the kwink bands is related to the dislocation slip, which is an energy dissipating process, and the dissipation may prevent the system from reaching the exact energy-minimizing configuration. Again, such an effect cannot be expected for classical twinning. 

Let us point out that the misorientation for the $(20\bar{1})_{\rm M}$ twins are then also inherited by the $(41\bar{1})_{\rm P}$ twins in the parent phase after the reverse transformation; a large scatter of the observed orientations from those predicted theoretically was reported in \citep{Molnarova_2020}. This agrees well with the fact that the $(41\bar{1})_{\rm P}$-twinning mechanism in the B2 structure is also quite energetically demanding and including a large-amplidute shuffle \citep{Christian_Acta_1988}; hence, as an alternative mechanism we suggest that the $(41\bar{1})_{\rm P}$ twins are, in fact, not twins but symmetric-tilt grain boundaries that are obtained from kwinks formed in martensite.

\subsubsection{Plastic forming of oriented martensite}

The second requirement we put on the model is that it must be able to capture the inelastic straining of the B19$^\prime$ lattice in case of fully detwinned oriented martensite, loaded along the $[10\bar{1}]_{\rm M}$ direction, which is the direction of the largest transformation strain in the $(010)_{\rm M}$ plane. This situation appears for example at the yielding plateu in cold-worked NiTi wires reported in \citep{Molnarova_2020} in grains with the vector ${\mathbf p}$ pointing along $[10\bar{1}]_{\rm M}$.  This loading direction corresponds to the $[111]_{\rm P}$ direction in austenite, which means the dominant axial direction in the textured wire. From a more general point of view, this is the most fundamental requirement to be put on any model of plastic forming of NiTi martensite: it should be able to identify what mechanisms are activated once all reversible twinning systems are exhausted.  

\begin{figure}
 \centering
 \includegraphics[width=0.9\textwidth]{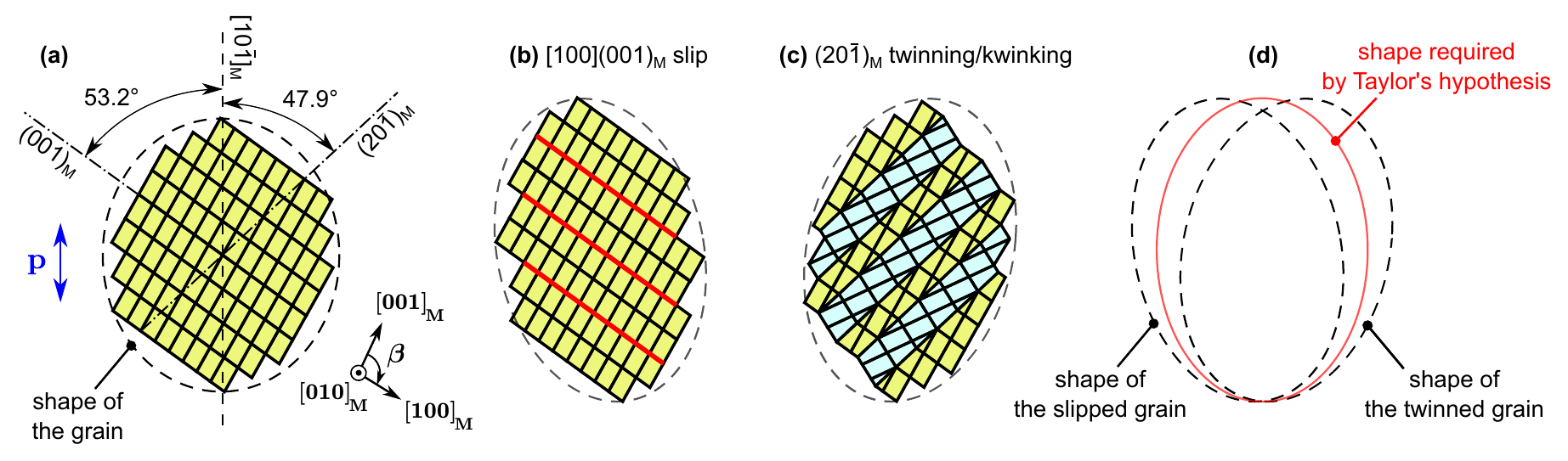}
 \caption{Visualization of plastic forming of a grain consisting only of fully oriented, detwinned martensite: (a) a single-variant grain loaded along the direction of maximum transformation strain in the given $(010)_{\rm M}$ plane, the dash-dot lines show the traces of invariant planes for the slip and twinning/kwinking, ${\mathbf p}$ is the projection of the wire axis onto the given $(010)_{\rm M}$ plane; (b) the change of the shape of the grain induced by the $[100](001)_{\rm M}$-dislocation slip, the red lines denote the planes at which the slip occurred; (c) the change of the shape of the grain induced by the $(20\bar{1})_{\rm M}$ plastic twinning (or kwinking); (d) comparison of the resulting shapes of the grain from (b) and (c) with the shape representing area-preserving elongation along the vector ${\mathbf p}$.}\label{plastic}
\end{figure}

We consider a single variant of martensite subjected to tensile loading along $[10\bar{1}]_{\rm M}$ (Figure \ref{plastic}(a)). The $(001)_{\rm M}$ and $(100)_{\rm M}$ twins have negative Schmid factors, i.e., they cannot contribute to any further elongation of the lattice along the loading direction. Hence, the only available deformation mechanisms with positive Schmid factors are the $[100](001)_{\rm M}$ dislocation slip (possibly leading to formation of kink bands) and the $(20\bar{1})_{\rm M}$ plastic twinning (or kwinking, as introduced above).

As outlined in Figures \ref{plastic}(b) and (c), respectively, both these mechanisms provide elongation along the $[10\bar{1}]_{\rm M}$ direction, but also a shear deformation in a perpendicular direction; the perpendicular shear deformation is quite pronounced, since the angle between the invariant planes and the loading direction is larger than $\pi/4$ in both cases. 
According to the Taylor's hypothesis \citep{Bhattacharya_Acta_1996,Ono_JIM_1989,Shu_Acta_1998}, the shape change of individual grains in the polycrystal must comply with the macroscopic straining of the wire. This requirement follows from the fact that the grains are anchored in the microstructure, i.e., they cannot freely rotate  with respect to each other. In other words, the Taylor's hypothesis requires the grain to achieve pure elongation along the loading direction without the perpendicular shear strains, which may happen only if the
$[100](001)_{\rm M}$ dislocation slip and the $(20\bar{1})_{\rm M}$ twining/kwinking appears simultaneously, such that the perpendicular strains compensate (Figure \ref{plastic}(d)).  Hence, the $(20\bar{1})_{\rm M}$ twinning/kwinking in this case is triggered if and only if the lattice, at the same time, plastically slips along the $(001)_{\rm M}$ planes. Nearly exact compensation of the shears is possible, since the invariant planes for both mechanisms make quite similar angles with the loading axis (Figure \ref{plastic}(a)).

At the yielding plateau (Figure \ref{evolution}(e)), the $(20\bar{1})_{\rm M}$ bands continuously grow, which means that also the whole lattice continuously slips such that the shears in the perpendicular direction are compensated. This is an additional motivation for not considering the $(20\bar{1})_{\rm M}$ plastic twinning as an independent mechanism from the dislocations slip. The concept of kwinking assumes, instead, that the $(20\bar{1})_{\rm M}$ bands are direct consequences of the slip, and thus, the simultaneous $(20\bar{1})_{\rm M}$ plastic twinning and plastic slip are, in fact, one straining mechanism, with the  $[100](001)_{\rm M}$ dislocations moving everywhere in the material: homogeneously in the matrix and inside the bands, the motion of the kwink (or kink) interfaces between the matrix and the bands itself can be also seen as a coordinated motion of slip dislocations. The kwink interfaces are composed of dislocation cores, and they are, therefore, probably prone to absorbing or releasing more dislocations, enabling the plastic slip to take place on both sides of the interface. In contrast, the twin interfaces typically act as obstacles against the dislocation slip. Hence, again, we can conclude that the desired feature of the model is more easily achieved when the $(20\bar{1})_{\rm M}$ are considered to be kwink interfaces instead of twin interfaces.

In Section 4.2, we will show by an explicit calculation that there exist kwink bands that enable an exact compensation of the perpendicular shears from the $[100](001)_{\rm M}$ slip, and that these bands are, again, oriented close to the $(20\bar{1})_{\rm M}$ planes. In some sense, this observation is quite similar to the observation done in the previous subsection, where the approximately $(20\bar{1})_{\rm M}$-oriented kwink bands were shown to be able to cross compatibly with the $(001)_{\rm M}$ twin bands; here we assume that they can also co-exist with the $[100](001)_{\rm M}$ slip that carries shear strains of the same orientation as the $(001)_{\rm M}$ twinning.

\bigskip

\subsubsection{V-shaped patterns and symmetric-tilt grain boundaries}
The last heuristic argument we use for justification of the proposed model is that this model is able to fully capture the pattern formation observed in the experiments in the given $(010)_{\rm M}$ martensite plane, and to explain it in terms of energy reduction. The TEM observations by  \cite{Molnarova_2020} prove that there are no other patterns created at the yielding plateau than the bands and V-shaped microstructures in the $(010)_{\rm M}$ plane. And thus, if the model can capture them, it can fully capture the whole plastic forming process.

\begin{figure}
 \centering
 \includegraphics[width=\textwidth]{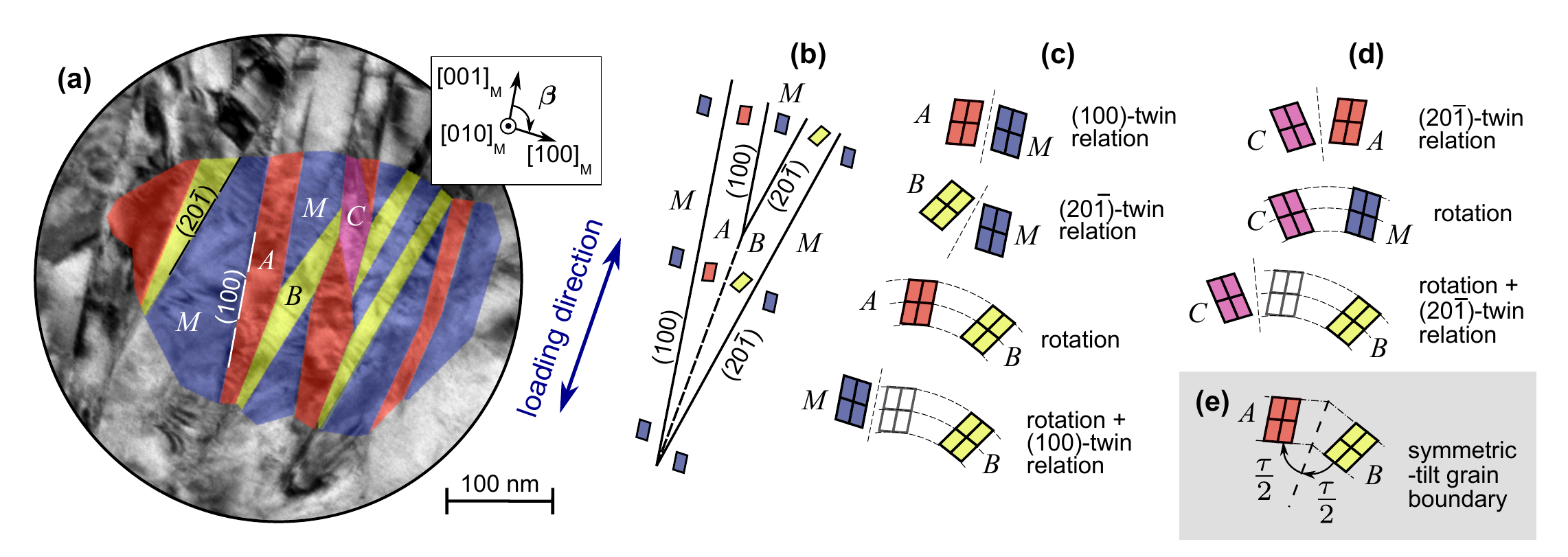}
 \caption{(a) TEM image of a representative grain in plastically formed NiTi; (b) notation and the spatial arrangement of the three lattices ($M$, $A$, and $B$) forming the V-shaped pattern; (c) orientation relationships within the pattern; (d) orientation relationships for the secondary $(20\bar{1})_{\rm M}$ twinning (lattice $C$); (e) mirror symmetry of the $(001)_{\rm M}$ slip planes with respect to the $A$-$B$ interface, with $\tau$ representing the tilt angle. The shapes of the unit cells as well as the orientations of the assumed twin interfaces in (b) correspond to the lattice parameters from \citep{Bhattacharya}, the orientation of the $A$-$B$ interface with respect to the twin interfaces and lattice orientations for (b) and (e) is taken from the micrograph in (a).  The loading direction (or, more precisely, its projection onto the given $(010)_{\rm M}$ plane) is approximately perpendicular to the $(001)_{\rm M}$ planes in lattice $M$.}\label{wedge}
\end{figure}

We focus on grains that are loaded in direction approximately perpendicular to $(001)_{\rm M}$, because of the richer patterns appearing in them. For the argument, we use the microstructure in one selected grain from \citep{Molnarova_2020}, containing prototypical V-shaped patterns and secondary  $(20\bar{1})_{\rm M}$ twinning. The grain is shown in Figure \ref{wedge}(a) (see Figure 10 in \citep{Molnarova_2020} for the details of the indexing of individual lattices forming the pattern). The grain is composed of the matrix lattice ($M$) and lattices ($A$) and ($B$) holding $(100)_{\rm M}$ and $(20\bar{1})_{\rm M}$ twin relationships to the matrix lattice, respectively. The both twin relationships are of the compound character, which means that the lattice $A$ is mirror-symmetric to $M$ over the $(100)_{\rm M}$ plane, and so is the lattice $B$ with respect to $M$ over the $(20\bar{1})_{\rm M}$ plane. The arrangement of these lattices in a simple V-shaped microstructure is visualized in Figure \ref{wedge}(b). In addition, there is the lattice $C$, which holds a $(20\bar{1})_{\rm M}$ twin relationship with the lattice $B$, that is, $C$ is a result of the secondary $(20\bar{1})_{\rm M}$ twinning, reported already in \citep{Zhang}.

Let us now analyze the characters of the individual interfaces between the lattices $M$, $A$, $B$, and $C$, and discuss whether they can arise from the simple mechanisms we build our model from, which are the martensite reorientation through reversible $(100)_{\rm M}$ and $(001)_{\rm M}$ twinning and the $[100](001)_{\rm M}$ dislocation slip.
For the primary twinning, this analysis is visualized in Figure \ref{wedge}(c). The $A$-$M$ twinning is directly one of these considered mechanisms. Based on the concept of kwinking, the $B$-$M$ relation (i.e. the $(20\bar{1})_{\rm M}$ twin relation) can be decomposed into two steps: one is a simple rotation in the $(010)_{\rm M}$ plane, the second is the $(100)_{\rm M}$ twinning. Importantly, the rotation angle is the same as the rotation relating the lattices $A$ and $B$, which is a necessary condition for the whole microstructure being geometrically admissible. For the secondary twinning (Figure \ref{wedge}(d)), similar decompositions can be done for relationships between the lattice $C$ and all other lattices. The $C$-$A$ relationship is the same as $B$-$M$, which has been discussed above. The $C$-$M$ relationship is a simple rotation, which can be understood as merging the $C$-$A$ twinning and $A$-$M$ twinning, and finally the $C$-$B$ relationship can be decomposed into the $C$-$A$ relationship and a rotation, again identical to the rotation between $A$ and $B$. In summary, each lattice from the observed pattern can be translated into any other lattice from the pattern by a sequence of two mechanisms: $(100)_{\rm M}$ twinning and a specific rotation, equivalent the rotation between lattices $A$ and $B$.

To confirm that the formation of the observed pattern can be indeed explained by the proposed model, we need to show that this specific rotation can be a result of the $[100](001)_{\rm M}$ dislocation slip. As it can be concluded both from crystallographic considerations \citep{Starkey_Mineral_1968} and from continuum-mechanics compatibility conditions \citep{Inamura}, the kink interface between two mutually rotated lattices resulting from unidirectional dislocation slip must always be a mirror plane between orientations of the slip planes on both sides of the kink interface. This follows from a simple requirement that the interface must be equally distorted and rotated in both lattices. The interfaces satisfying this condition are the so-called \emph{symmetric-tilt grain boundaries} (STGB, see e.g. \citep{Han_Acta_2016}), that can be characterized by a single scalar parameter, the tilt angle $\tau$. For the needs of the model formulated in Section 3, we define this angle as the angle between the slip planes in the neighboring lattices, where for a STGB both slip planes make an angle of $\tau/2$ with the interface. Notice that for low-symmetry lattices, such as the monoclinic lattice of B19$^\prime$, the mirror-symmetry for STGBs holds only for the slip plane orientations, not for the lattices themselves.  This distinguishes here the STGBs from the twins.

In our case, this condition means that the $A$-$B$ interface should be a mirror-symmetry plane between $(001)_{\rm M}$ planes in lattices $A$ and $B$. Figure \ref{wedge}(e) shows that, within the precision of $\pm{}2^\circ$ with which the orientation of the interface can be assessed from the micrograph, there is a perfect mirror symmetry between the $(001)_{\rm M}$ planes, with $\tau\approx{}146^\circ$. And thus, indeed, the $A$-$B$ rotation for the given orientation of the $A$-$B$ can arise as a result of the $[100](001)_{\rm M}$ dislocation slip. We will further support this conclusion by an explicit calculation in Section 4.3. For the needs of the current heuristic discussion, nevertheless, we can conclude that the $A$-$B$ interfaces can be captured by the proposed model, and so can also all other interfaces observed in the pattern, because (as it can be easily checked) all of them satisfy the requirement of the mirror symmetry between the slip planes. In other words, all these interfaces are admissible as the outputs of the model.

It is also worth noting that this may not be true for all planar interfaces found in martensitic microstructures. \cite{Inamura_Acta_2017} observed interfaces created by head collisions of mutually incompatible martensitic plates in a Ti-Nb-Al alloy formed within the phase transition from austenite. Although being visually similar to the V-shaped patterns in NiTi, these microstructures were of a very different origin and behavior. While the V-shaped patterns gradually grow during the plastic forming, exploiting the twin/kink/kwink character of all interfaces, the martensite plates in \citep{Inamura_Acta_2017} stop growing along the junction plane immediately as they collide, which results in fine triangular microstructures. The reason is that the collision-created interfaces violate the compatibility conditions, and are, thus, energetically expensive.   

\bigskip
Finally, we need to suggest how the formation of the V-shaped patterns can lead to low-energy accommodation of the tensile strains applied to the grain. Here we utilize the knowledge of the deformation mechanisms in mechanical systems with unidirectional slip \citep{Lei, LPSO, Inamura, Wadee}. As shown above, all interfaces present in the microstructure are either twin interfaces or kink interfaces or their combinations, and, as such, they can be expected to appear in energy-minimizing patterns.
This does not, however, guarantee that the whole pattern can be also an energy minimizer or even its close approximation. The interfaces meet in triple junctions, where certain geometric conditions between the shearing directions \citep{Bhattacharya_Acta_1991,Bhattacharya,Balandraud_JMPS_2007} need to be satisfied. If they are not, a stress singularity (a so-called disclination \citep{Inamura,Mullner_MSEA_1997,Mullner_Acta_2000,Mullner_Acta_2010}) arises at the junction. This would lead to a tendency to merge pairs of triple junctions together such that the disclinations get compensated \citep{Inamura}, which has not been ever reported from the experiments on NiTi martensite (cf. \citep{Zhang, Molnarova_2020}). 

\begin{figure}
 \centering
 \includegraphics[width=0.8\textwidth]{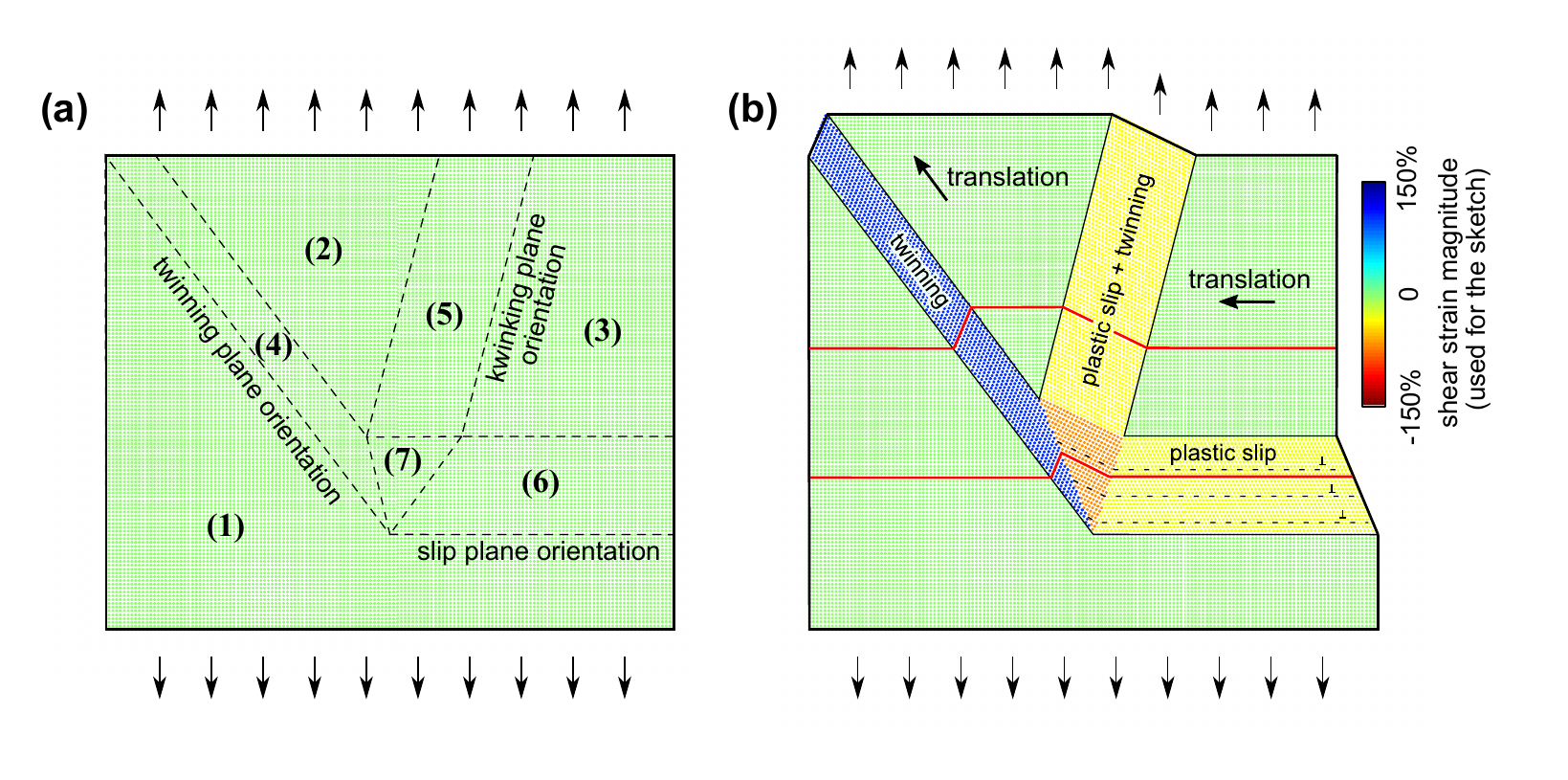}
 \caption{A schematic sketch of the pattern formation mechanism: (a) reference configuration and (b) deformed configuration of a representative volume accommodating tensile loads in the vertical direction by formation of a V-shaped microstructure. In (a), the numbers (1)-(7) denote regions gaining different homogeneous deformation gradients and/or translations in the deformed configuration. The color code in (b) represents the magnitude of shear along the planes that are horizontal in (a), that is, the slip planes. The red lines represent two such planes in the deformed configuration; for a real material, these planes would be $(001)_{\rm M}$ planes along which the slip in the $[100]_{\rm M}$ direction occurs. It is seen that these planes are mirror-symmetric with respect to each interface they intersect. {{}While in real V-shaped patterns in B19$^\prime$ martensite the shear strain magnitude in the $(100)_{\rm M}$ twin bands is $\varepsilon^{\rm twin} =0.2385$ (see Table \ref{tabul}), for this sketch a more than six-times higher amplitude $\varepsilon^{\rm twin}=1.5$ is used for better visibility of the elongation of the domain along the loading direction.}}\label{cartoon}
\end{figure}

We will show in Section 4.3 that the conditions for the shearing vectors cannot be satisfied, regardless of the exact orientation of the involved twin and kwink (or kink) bands. Hence, there must be an additional mechanism, not visible from the micrographs, that compensates the stress singularity at the disclinations and enables, thus, the V-shaped patterns to form without any excessive increase in elastic energy in the grain. One possibility is that the disclination is compensated by the $[100](001)_{\rm M}$ plastic slip. A simple mechanism how this can be achieved is schematically sketched in Figure \ref{cartoon} (notice that for better visibility of the strains in this sketch, the used geometry is not derived from real lattice parameters, unlike in most of the visualizations in this paper). In this figure, we consider first the matrix lattice as the reference (undeformed) configuration, and we suggest a pattern composed of seven regions, neighboring to each other over planar interfaces (Figure \ref{cartoon}(a)). There are no triple junctions in this construction, but quadruple junctions, for which the compatibility conditions are easier to achieve\footnote{A somehow simplified reasoning is that a triple junction can exist without a disclination only if the meeting regions are all sheared in the direction parallel to the junction line only. While this is possible in some wedge-shaped martensitic microstructures \citep{Bhattacharya_Acta_1991}, it cannot be achieved here, because we assume that all involved mechanisms act only on the $(010)_{\rm M}$ plane. For the quadruple junctions, the condition is softer, as the four meeting deformation gradients can pairwise compensate.} \citep{Bhattacharya,JMPS}; this is reached by adding an extra band (regions (6) and (7)) to the V-shaped geometry.

In the deformed configuration (Figure \ref{cartoon}(b)), the regions (1), (2) and (3) remain in the matrix lattice, while the other gain homogeneous deformation gradients such that the continuity of the domain as a whole remains preserved. All involved deformation gradients can be polar-decomposed into a shape strain in the shown plane (the shape strain here is pure shear and the amplitude of this shear is used for the colors in Figure \ref{cartoon}(b)) and a rotation in this plane. The bands (4) and (5) become the twin band and the kwink band, respectively, with opposite signs of the shear deformation. The plastically slipped band (6) is oriented parallel to the slip planes, and that is why there are no visible interfaces between this band and the regions (1) and (3) in the micrograph. The role of the band (6) is mainly in bringing extra plastic strain into the triangular region (7), such that the disclination is fully compensated. As a result, the dislocation-induced shear strain in this triangle is of a higher amplitude than in the rest of the kwink band (5), but because the boundary between (7) and (5) is parallel to the slip plane, it is invisible in the micrograph. The dislocation cores that move though the band (6) might originate either from a grain boundary, or possibly from another V-shaped microstructure, oriented upside-down with respect to the discussed one (see \citep{Inamura} for a similar situation for kink ridge patterns). These dislocation cores enter the region (7) and stop at its boundary with the twin band (4), where contribute to the rotation between the adjacent lattices of the same variant of martensite. 

To confirm that the construction in Figure \ref{cartoon}(a) enables creation of the V-shaped microstructures appearing in the experiments, we need to show that the interfaces (6)-to-(7) and (3)-to-(5), although not parallel in the reference configuration (Figure \ref{cartoon}(a)), become always perfectly parallel to each other in the deformed configuration (Figure \ref{cartoon}(b).  This is done in Section 4.3, where also the explicit amount of dislocation slip in the band (6) needed to compensate the disclination is calculated for B19$^\prime$ NiTi martensite. From Figure \ref{cartoon}(b), it is also seen how the V-shaped pattern helps to accommodate tensile strains along the loading direction. The region (2) is translated upwards, along the twin boundary interface, which is enabled by the horizontal translation of the region (3). Such a mechanism, involving twinning, kwinking and plastic slip, results in elongating the considered domain, although the loading is applied perpendicular to the slip plane. This explains why the V-shaped patterns are seen mainly in grains with the $(001)_{\rm M}$ plane oriented perpendicular to the loading direction in nanocrystalline textured NiTi wires (see Section 2.2.2). In grains oriented such that the loading is applied along the $[10\bar{1}]$ direction, the combination of dislocation slip and kwinking is sufficient for accommodating the strains, as described in the previous Section 2.3.2.. Consequently, only the kwink bands appear.

\subsection{Summary of the heuristic argument}
We have above discussed three particular experimentally observed processes occurring in plastically strained NiTi, and in all cases we realized that these processes require the $[100](001)_{\rm M}$ dislocation slip to take place. Once the slip is active, the lattice can locally easily translate between different EPNs, and so distinguishing between 'twinning inside one EPN' and 'twinning beyond the EPN' becomes not important. As a substitute for the concept of 'twinning outside the EPN' we proposed the new concept of kwink bands, which are twin bands with additional rotations coming from the dislocation cores present at the interfaces{{}, and additional shear coming from the dislocation slip in the adjacent variants of martensite.} Approximately $(20\bar{1})_{\rm M}$-oriented kwinks appear to be of high importance: they can cross the $(00{1})_{\rm M}$-compound twins without additional elastic energy, which enables their easy nucleation beyond the end of the reorientation plateau, and they can also compensate the perpendicular shears coming from the $[100](001)_{\rm M}$ dislocation slip in oriented martensite, which enables them to act as a part of the mechanism of plastic forming at the yielding plateau (we support both these claims by direct calculations in Section 4). Finally, the $(20\bar{1})_{\rm M}$-oriented kwinks can compose into V-shaped patterns with $(100)_{\rm M}$ twins, provided that there is additional dislocation slip available, enabling compensation of the disclinations at the triple points. These heuristic results serve as a basis for the model developed in the following sections, where we show the appearance of specifically oriented kwinks and their co-existence with other twins and with massive dislocation slip can be understood as a result of energy minimization.

\section{Model formulation}

For the continuum-level description of the plastic forming mechanisms, we use the framework and formalism developed for martensitic microstructures within the theory of non-linear elasticity \citep{JMB1,JMB2,Bhattacharya}. In this theory, the individual variants of martensite are represented by deformation gradients constructed from the lattice correspondence and lattice parameters \emph{via} the Cauchy-Born hypothesis. The same formalism can be also applied for plastic slip, as utilized in crystal plasticity \citep{Roters_Acta_2010,Gurtin_JMPS_2000,Acharya_JMPS_2000}. As  discussed in detail in Section 2, the plastically formed  B19$^\prime$ NiTi martensite undergoes mainly $(100)_{\rm M}$, $(001)_{\rm M}$ and $(20\bar{1})_{\rm M}$ twinning and the $[100](001)_{\rm M}$ dislocation slip (see Table \ref{tabul}). If we restrict the construction of the model just to these mechanisms, the problem becomes essentially two-dimensional, as all these mechanism represent straining just in one $(010)_{\rm M}$ plane (one $\{110\}_{\rm P}$ plane in the parent phase), shared by all involved variants. This significantly simplifies the description.

\subsection{Variants of martensite and deformation mechanisms}

\begin{figure} 
 \centering
 \includegraphics[width=\textwidth]{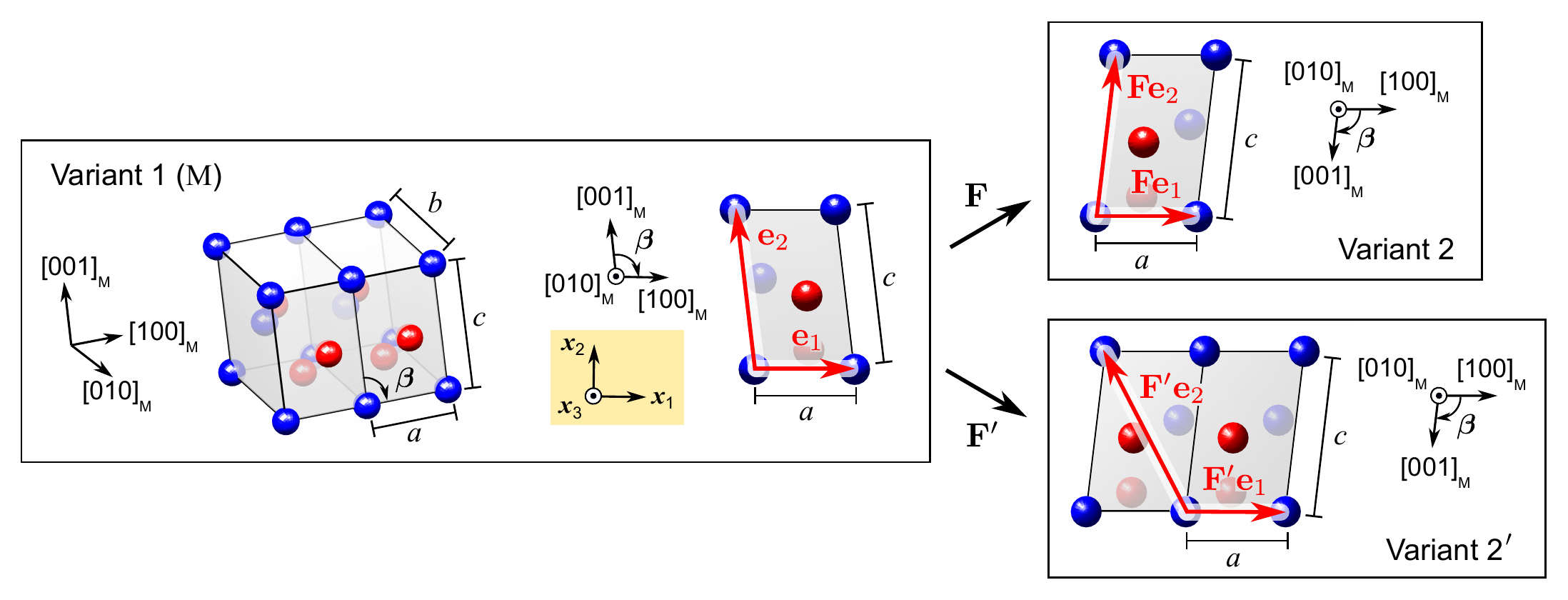}
 \caption{Notation of the variants, the used coordinate system ($x_i$ axes in the shaded bar), and the definition of the deformation gradients ${\mathbf F}$ and ${\mathbf F}^\prime$. The shapes of the unit cells correspond to the lattice parameters of B19$^\prime$ martensite from \citep{Bhattacharya}.}\label{cauchyborn}
\end{figure}

We take one variant of martensite as the reference configuration (Variant 1 in Figure \ref{cauchyborn}), let the lattice in the $(010)_{\rm M}$ plane be generated by vectors (${\mathbf e}_1$ and ${\mathbf e}_2$). We refer further to this variant as to the matrix ($M$) and assume that this variant is dominant in the given grain. The deformation gradient representing this variant is the identity (${\mathbf I}$), i.e. we fix the rotation of this variant. The matrix sets the coordinate system $x_i$ in which we describe the evolution of the microstructure,  using two-dimensional deformation gradients acting on the $x_1x_2$ plane,  see Figure \ref{cauchyborn}. 

The second variant sharing the same $(010)_{\rm M}$ plane with the matrix is the Variant 2 (on the right in Figure \ref{cauchyborn}), that can be achieved from Variant 1 by shearing. Two different shearing paths between Variant 1 and Variant 2 can be considered. The first is through the deformation gradient ${\mathbf F}$  and keeps the lattice in the given EPN. This path corresponds to the classical compound twinning, as explained below. The second path is through the deformation gradient ${\mathbf F}^\prime$  which goes beyond the given EPN. This path (suggested by  \cite{Gao,pathways}) can represent either the irreversible twinning, or the $[100](001)_{\rm M}$ dislocation slip of Variant 2 with a specific slip amplitude. To highlight that this path results in the same variant of martensite but corresponding to another EPN, we use the denotation Variant 2$^\prime$ in Figure \ref{cauchyborn}.

In the given coordinate system, the deformation gradients are 
\begin{equation}
 {\mathbf F} = \left(\begin{array}{cc}
                      1 & \gamma \\ 0 & 1
    \end{array}\right) {\mbox{\hspace{1cm}and\hspace{1cm}}}{\mathbf F}^\prime = \left(\begin{array}{cc}
                      1 & \gamma^\prime \\ 0 & 1
    \end{array}\right),\label{FFprime}
\end{equation}
where $\gamma=0.2385$ and $\gamma^\prime=-0.3910$ for the lattice parameters of monoclinic B19$^\prime$ martensite introduced in Section 2.1. The lattice of Variant 2 (or Variant 2$^\prime$) is generated by vectors ${\mathbf F}{\mathbf e}_1$ and ${\mathbf F}{\mathbf e}_2$ (or ${\mathbf F}^\prime{\mathbf e}_1$ and ${\mathbf F}^\prime{\mathbf e}_2$). Notice, however, that for Variant 2$^\prime$ the deformation gradient ${\mathbf F}^\prime$ describes only the change of the shape of the unit cell, not the rearrangement of the atoms inside it. In other words, the reorientation of martensite outside the given EPN requires also additional shuffling, as discussed in more details in Section 2.3. 

\bigskip
In addition to the deformation gradients representing the twinning, we also consider a deformation gradient representing the slip. Here we adopt the assumption of \cite{Kudoh_1985} that the easiest slip in the B19$^\prime$ martensite lattice is the $[100](001)_{\rm M}$ slip. According to the first-principles calculations \citep{Ezaz_MSEA_2020}, the energy barriers for activation  of this slip are comparable to those for all considered twins, and by an order of magnitude lower than those for the hypothetical $[001](100)_{\rm M}$ or $[\bar{1}0\bar{2}](20\bar{1})_{\rm M}$ dislocation slips.  In the given coordinate system, the two-dimensional deformation gradient representing the $[100](001)_{\rm M}$ slip in Variant 1 can be written as
\begin{equation}
 {\mathbf P}(\alpha) = \left(\begin{array}{cc}
                      1 & \alpha\\ 0 & 1
                     \end{array} \right),
\end{equation}
where $\alpha$ is the slip amplitude representing the average amount of slip in the given volume. Notice that there are specific slip amplitudes $\alpha=\gamma$ and $\alpha=\gamma^\prime$ for which ${\mathbf P}(\gamma)={\mathbf F}$ and ${\mathbf P}(\gamma^\prime)={\mathbf F}^\prime$, i.e. the same effective deformation of the lattice can be achieved by slip of Variant 1, and by reorientation of Variant 1 into Variant 2 or Variant 2$^\prime$. 

Plastic slip can be further combined with martensite reorientation, resulting in the deformation gradient  \begin{equation}
{\mathbf P}(\alpha){\mathbf F}={\mathbf F}{\mathbf P}(\alpha),\label{commute}                                                                                            
\end{equation}
and equivalently for ${\mathbf F}^\prime$. Note that the commutability follows from the specific form of ${\mathbf P}(\alpha)$, ${\mathbf F}$, and ${\mathbf F}^\prime$. 

\subsection{{}Compatibility conditions and coherent interfaces}
When a microstructure is formed, different regions of the material undergo different deformations. To achieve the kinematic compatibility (i.e., displacement field continuity) between these regions, additional rotations may be needed. For this reason, we assume henceforth the deformation gradients of form ${\mathbf R}{\mathbf F}$, ${\mathbf R}{\mathbf F}^\prime$ and ${\mathbf R}{\mathbf P}(\alpha)$ as those of which the analyzed patterns are composed, where ${\mathbf R}\in SO(2)$ is the orthogonal rotation matrix. In the given two-dimensional setting, the rotation can be represented by one angle $\theta$, with 
\begin{equation}
{\mathbf R}(\theta)= \left(\begin{array}{cc}
                      \cos{\theta} & -\sin{\theta}\\ \sin{\theta} & \cos{\theta}
                     \end{array} \right). \label{rotace}
\end{equation}

The kinematically compatible connection between two deformation gradients ${\mathbf G}_1$ and ${\mathbf G}_2$ is possible 
if there exist a rotation matrix ${\mathbf R}$ and the vectors ${\mathbf m}$  ($|{\mathbf m}|=1$) and ${\mathbf b}$ ($|{\mathbf b}|>0$), such that
\begin{equation}
 {\mathbf R}{\mathbf G}_2 - {\mathbf G}_1 = {\mathbf b} \otimes {\mathbf m}. \label{compat_G}
\end{equation}
Then, there can exist a planar coherent interface between homogeneously deformed regions with deformation gradients ${\mathbf R}{\mathbf G}_2$ and ${\mathbf G}_1$, the vector ${\mathbf m}$ is perpendicular to this interface, and the vector ${\mathbf b}$ (the shearing vector) gives the strain jump over the interface.

\begin{figure}
 \centering
 \includegraphics[width=\textwidth]{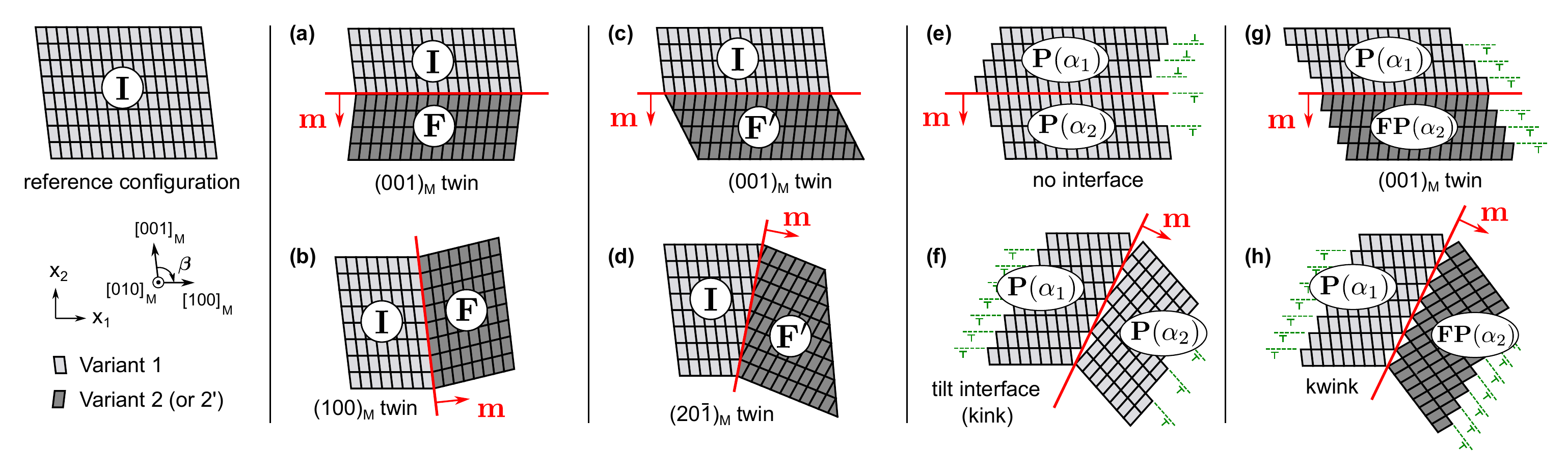}
 \caption{Possible coherent interfaces created between lattices with deformation gradients ${\mathbf I}$, ${\mathbf F}$, ${\mathbf F}^\prime$, ${\mathbf P}(\alpha)$ and their (commutable) products. The dashed lines with dislocation signs outline the orientations of the $[100](001)_{\rm M}$ slip planes with respect to the interface. ${\mathbf m}$ in each drawing denotes the unit vector perpendicular to the coherent interface, solving the equation (\ref{compat_G}).}\label{coherent}
\end{figure}

Depending on the character of the deformation gradients ${\mathbf G}_1$ and ${\mathbf G}_2$, we distinguish henceforth between three different types of compatible interfaces that can arise in the discussed microstructures:

\begin{enumerate}
 \item if ${\mathbf G}_1={\mathbf I}$ and ${\mathbf G}_2\in\left\{{\mathbf F},{\mathbf F}^\prime\right\}$, the interface is a twin interface. Two such parallel interfaces encapsulate \emph{a twin}, i.e., a band of Variant 2 (or Variant 2$^\prime$) inside the matrix. In particular, if ${\mathbf G}_2={\mathbf F}$, the equation (\ref{compat_G}) has two solutions. One solution is  $\theta=0$, ${\mathbf b}=(\gamma; 0)$, ${\mathbf m}=(0; -1)$, which represents $(001)_{\rm M}$ compound twinning (Figure \ref{coherent}(a)). The second solution is $\theta=13.6^\circ$, ${\mathbf b}=(-0.0282; 0.2368)$, ${\mathbf m}=(0.9930; 0.1184)$, which represents the $(100)_{\rm M}$ twinning (Figure \ref{coherent}(b)).  When  ${\mathbf G}_2={\mathbf F}^\prime$, one solution of (\ref{compat_G}) is again with $\theta = 0$ and corresponds to a $(001)_{\rm M}$ twin, in which, however, the matrix and the twin correspond to a different EPN (Figure \ref{coherent}(c)). The second solution is the $(20\bar{1})_{\rm M}$ twinning with parameters $\theta=-22.12^\circ$, ${\mathbf b}=(-0.0750; -0.3837)$, ${\mathbf m}=(0.9814;-0.1919)$ (Figure \ref{coherent}(d)).
 
 \item if ${\mathbf G}_1={\mathbf P}(\alpha_1)$ and ${\mathbf G}_2={\mathbf P}(\alpha_2)$, where $\alpha_1\neq\alpha_2$ are two different slip amplitudes, one solution is trivial with $\theta=0$, ${\mathbf m}=(0; -1)$ and ${\mathbf b}$ given by the difference between $\alpha_1$ and $\alpha_2$ (Figure \ref{coherent}(e)). The second solution has $\theta \neq{}0$ and represents a tilt interface, i.e., an interface  composed of dislocation cores providing a mutual rotation between two slipped lattices (Figure \ref{coherent}(f)). The tilt across the interface is given by the average spacing between the dislocation cores. Such interfaces typically appear in materials with highly anisotropic plastic slip, such as low-period stacking-order (LPSO) magnesium alloys \citep{LPSO,Lei,Yamasaki,Takagi}, or in layered media \citep{Wadee}. A band encapsulated by two tilt interfaces is usually called \emph{a kink}. Notice that unlike twins, the kinks are not directly related to specific crystallographic planes, as their orientation is determined by the jump of the slip magnitude $\alpha$ over them; they can form ridge-like patterns and even more complex microstructures. The only requirement on the crystallographic orientations for the tilt interfaces is that they are mirror-symmetry planes between the $(001)_{\rm M}$ slip planes in regions with deformation gradients ${\mathbf G}_1$ and ${\mathbf G}_2$, that is, the tilt interfaces are \emph{symmetric-tilt grain boundaries} (STGBs), as discussed in Section 2.3.4. The tilt interfaces (and the kinks) can appear also in Variant 2, when ${\mathbf G}_1={\mathbf F}{\mathbf P}(\alpha_1)$ and ${\mathbf G}_2={\mathbf F}{\mathbf P}(\alpha_2)$. 
 \item{} if ${\mathbf G}_1={\mathbf P}(\alpha_1)$ and ${\mathbf G}_2={\mathbf F}{\mathbf P}(\alpha_2)$ (or ${\mathbf G}_2={\mathbf F}^\prime{\mathbf P}(\alpha_2)$), one solution is, again, trivial and represents a $(001)_{\rm M}$ twin between two differently slipped lattices (Figure \ref{coherent}(g)). As the slip runs along the same plane as the twinning, these two deformation mechanisms do not interact with each other.  The second solution represents a tilt interface between two variants, i.e., a kwink (Figure \ref{coherent}(h)).  Similarly to the twins, the kwinks are interfaces between different variants of martensite, and, at the same time, the tilt between the lattices must always be symmetric (as it is in STGBs) to achieve the compatibility. Consequently, the two lattices forming the kwink are necessarily twin-related. As discussed below, for a specific slip-amplitude jump $\alpha_2-\alpha_1$, this twin relation can be a $(20\bar{1})_{\rm M}$ twin relation. That is, the geometric relationship between the lattices in Figure \ref{coherent}(d) can be obtained as a particular case of the situation sketched in Figure \ref{coherent}(h).    
\end{enumerate}

\subsection{Energy considerations}

For the needs of the model, we postulate that the evolution of the microstructure and the pattern formation are driven by the reduction of the total energy of the material. We restrict our considerations to energy reduction in one representative grain (i.e., a single crystal in the corresponding austenite lattice),  and assume  that the microstructure evolves similarly in most of the grains, which leads to energy reduction over the whole polycrystalline aggregate.  The microstructure in plastically formed B19$^\prime$ evolves under external stress, which means that the proper energy description should be through minimization of Gibbs free energy. This energy can be understood as consisting of two parts: The first one comes from the linearly-elastic response of the lattice to the external loads and corresponds to small ($\lesssim 1\%$) but energetically expensive strains. The second one comes from the evolution of the microstructure (twinning, dislocation slip), and corresponds to large ($\sim 10\%$) strains that can be, however, energetically very cheap, provided that the microstructure evolves such that the kinematic compatibility conditions are satisfied at all involved interfaces and their junctions and intersections. Treating both these parts simultaneously would be quite difficult. The elastic strains from the external loads change the lattice parameters, which affects the deformation gradients representing the twinning or the dislocation slip, and, at the same time, the microstructure locally changes the lattice by rotations and switching to other variants of martensite, which alters the local response of the lattice to the stress, especially since the linearly-elastic behavior of the shape memory single crystals is known to be strongly anisotropic \citep{Wagner}.

For these reasons, we choose a simplified approach. Instead of minimizing the Gibbs free energy under considered stress, we assume minimization of Helmholz free energy under prescribed strains, that is, strains that relax the external loading. For example, in a grain inside of a thin wire loaded in tension, the elongation along the loading direction can be (in the given two-dimensional approximation) achieved by triplets of bands of homogeneous shear strain, as in the scenario sketched in Figure \ref{cartoon}(b), where the bands are a twin, a kwink, and a band of localized plastic slip compensating the disclination, respectively. The deformed configuration in Figure \ref{cartoon}(b) is energetically cheap, because it is perfectly compatible without elastic strains. At the same time, it accomodates the strains imposed by the external loading. In this sense, the minimization of the total Gibbs energy under the prescribed external stress can be well approximated by constructing a microstructure that, under zero external stress, would result in the required change of the shape of the grain with minimal Helmholz free energy. We utilized this approach earlier in \citep{Molnarova_2020} for explaining the fine $(001)_{\rm M}$ twinning in textured wires, or in \citep{Frost_IJSS_2021} for rationalizing the anisotropic nucleation of stress-induced martensite in twisted tubes. In Section 4, we make several similar \emph{ad hoc} constructions of energy-reducing microstructures. In some sense, this is also somehow consistent with the experimental procedure, where the TEM foils are fabricated \emph{a posteriori} from the plastically strained material after the loading is removed.

\bigskip
The total Helmohlz free energy of a given grain can be expressed as a sum of the bulk energy ($W_{\rm bulk}$) and the surface energy of the interfaces inside the grain ($W_{\rm surf}$). This reads
\begin{equation}
 W=W_{\rm bulk} + W_{\rm surf} = \int_V\varphi{({\mathbf G})}{\rm d}V + \sum_i\Phi_i{}S_i, \label{totalE}
\end{equation}
where $\varphi{({\mathbf G})}$ is the bulk energy density that depends on the deformation gradient ${\mathbf G}$, and $S_i$ are areas of the individual interfaces, each having a surface energy density $\Phi_i$. We assume that the interfaces are approximately planar, each having a constant surface energy density over its whole area. Notice that here we cannot use the form of the surface energy utilized often in the literature concerned with martensitic microstructures, where the energy density is assumed to be proportional to the jump of the deformation gradient over the interfaces (see e.g. \citep{Bhattacharya} for a discussion and relevant references). Our model needs to treat three different types of interfaces, the twins, the tilt interfaces, and the kwinks, each having a different atomistic-scale morphology, and thus, different surface energy. Especially  the latter two may have a very non-trivial dependence of the surface energy on the deformation gradient jump, as discussed below.

\begin{figure}
 \centering
 \includegraphics[width=\textwidth]{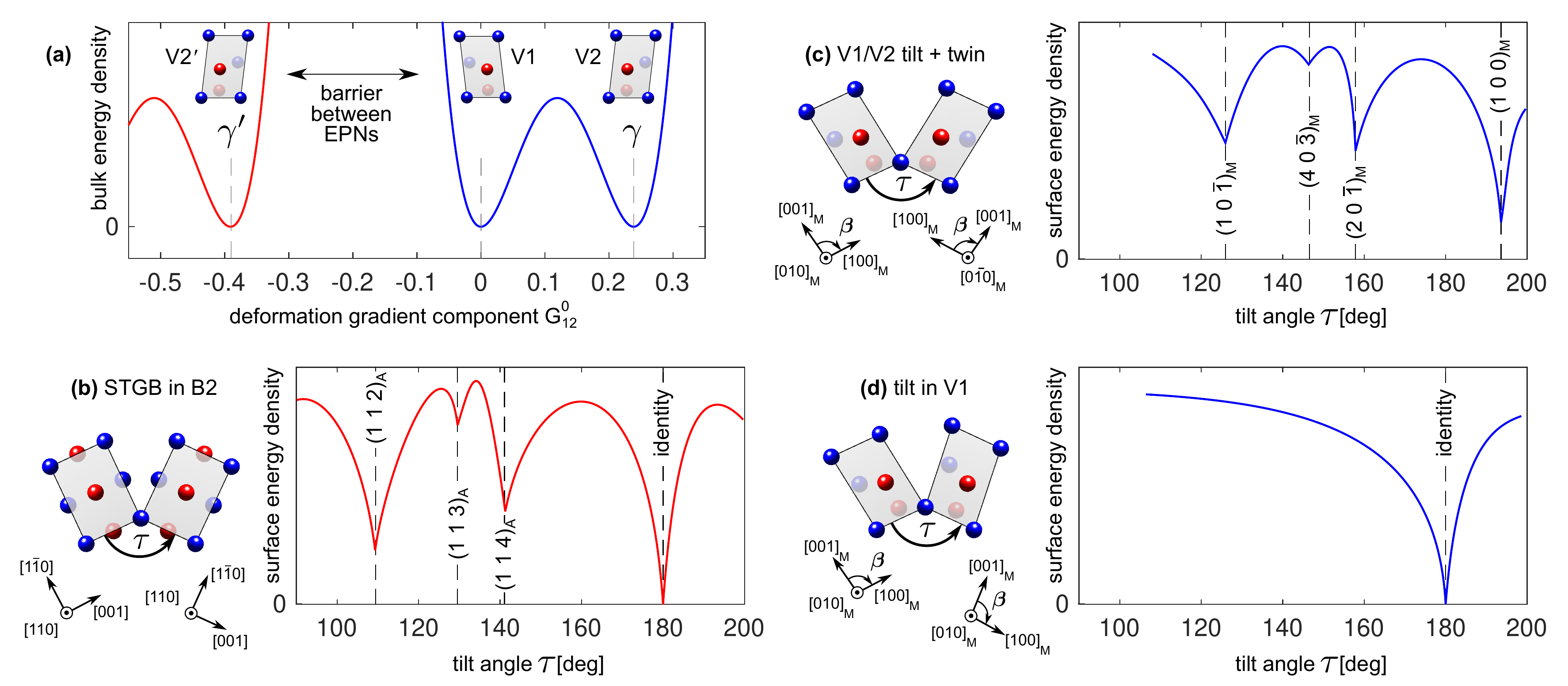}
 \caption{Assumed energy landscapes for the model: (a) bulk elastic energy ($W_{\rm bulk}$) density in the stress-free state, having a two-well character for each EPN; (b) symmetric-tilt grain-boundary surface energy for B2 austenite, the character of the landscape adopted from \citep{Yadzandoost_CompMaterSci_2017}; (c) a hypothetical symmetric-tilt grain-boundary surface energy for a kwink in B19$^\prime$, the character of the landscape obtained from (b) \emph{via} the lattice correspondence, with assuming non-zero twin energy for $\tau=180^\circ$; (d) symmetric-tilt grain-boundary surface energy for a tilt interface in  B19$^\prime$, no local minima appear at the curve as no low-$\Sigma$ states are achieved.  }\label{fig_energy}
\end{figure}

\bigskip
As introduced in the previous subsections, a stress-free deformation gradient resulting from the twinning and the dislocation slip can be always expressed as 
\begin{equation}
 {\mathbf G}={\mathbf R}{\mathbf G}^0{\mathbf P}(\alpha), \label{GRFP}
\end{equation}
where ${\mathbf R}$ is a rotation matrix,  ${\mathbf G}^0$ determines in which variant of martensite the material is, and ${\mathbf P}(\alpha)$ results from the dislocation slip. The bulk energy density is then affected only by the middle term, i.e., $\varphi({\mathbf G})=\varphi({\mathbf G}^0)$. The set ${\mathbf G}^0\in{}\left\{{\mathbf I},{\mathbf F},{\mathbf F}^\prime\right\}$ determines the minima of $\varphi({\mathbf G}^0)$; close to these minima the energy is expected to be convex, representing the elastic response of the given variant. The deformation gradients representing the variants 1, 2 and 2$^\prime$ differ just in the off-diagonal term $G^0_{12}$, which is equal to zero for Variant 1, and $G^0_{12}=\gamma$ or $G^0_{12}=\gamma^\prime$ for Variant 2 and Variant 2$^\prime$, respectively. Hence, we can assume that the bulk energy density has the character drawn in Figure \ref{fig_energy}(a) with respect to $G^0_{12}$. It has two minima corresponding to two variants of martensite, separated by an energy barrier which represents the energy required for $(001)_{\rm M}$ or $(100)_{\rm M}$ compound twinning. A much higher barrier can be assumed to exist between the minima in $G^0_{12}=0$ and $G^0_{12}=\gamma^\prime$, which are variants belonging to different EPNs. As suggested by \cite{Gao,pathways}, this barrier can be partially lowered due to an existence of a meta-stable intermediate structure of orthorhombic symmetry; as an alternative explanation why the material can overcome this barrier under high loads, we suggest here (see Section 2.3) that  Variant 2$^\prime$ is created from Variant 2 by an appropriate amount of plastic slip, which is \begin{equation}
  \alpha=-(\gamma-\gamma^\prime)=-0.6295,                                                                                                                                                                                                                                                                                                                                                                                                                                                                                                                                                                                                                                                                                                                                                                                                                                                                                                                                                                                                                                                                                                                                                                                                                                                                                                                                                                                                                                                                                                                    \end{equation}
  where $\gamma$ and $\gamma^\prime$ are those from definitions (\ref{FFprime}).  
This amount of slip is very similar to the slip required to form the typical wedge-shape patterns in LPSO magnesium alloys \citep{Inamura,Lei}, and microscopically corresponds to one $[100](001)_{\rm M}$ dislocation core on each $(001)_{\rm M}$ plane, {{} which means that the average distance between the dislocation cores at the perfect $(20\bar{1})_{\rm M}$ kwinking plane is essentially equal to the length of the lattice vector $\frac{1}{2}[102]_{\rm M}$}, see Figure \ref{kulicky} for a visualization. We assume that the slip does not increase the bulk energy but affects the total energy by the surface energy of the kwink or tilt interfaces that form as the consequence of the slip. In reality, the dislocation slip is associated with energy dissipation, i.e., for simulations of the time evolution of the microstructures, the derivative $\dot{\alpha}$ might be an important parameter. Such an extension of the model, however, falls beyond the scope of the paper. 

Because of assuming that Variant 2$^\prime$ is always achieved by the plastic slip of Variant 2, and, in a more general sense, that all twinning beyond the EPN is always created by an appropriate combination of twinning inside the EPN and plastic slip, the bulk energy $W_{\rm bulk}$ can be henceforth considered as a simple two-well potential with minima in $G^0_{12}=0$ and $G^0_{12}=\gamma$. In other words, we further assume just ${\mathbf G}^0\in{}\left\{{\mathbf I},{\mathbf F}\right\}$ as the bulk-energy minimizing set, instead of  ${\mathbf G}^0\in{}\left\{{\mathbf I},{\mathbf F},{\mathbf F}^\prime\right\}$.

\bigskip
The surface energy $W_{\rm surf}$ is a sum of energies of planar interfaces. The energy density $\Phi_i$ at the given interface depends on the type of the interface and also on the orientation relationship between the two lattices meeting at the interface. The energy is reduced for special relationships, where the two lattices share a high number of atomic sites. Special cases are the $(001)_{\rm M}$ and $(100)_{\rm M}$ compound twins, where all atomic sites are shared by both lattices, and thus, the corresponding surface energy can be assumed as very low (the same can be deduced from atomistic simulations, \citep{Li_Acta_2020}). From the fact that the $(001)_{\rm M}$ laminates observed in the microstructure are typically very fine, while the $(100)_{\rm M}$ twin bands are much thicker, one can deduce that $\Phi^{(100)}\gg\Phi^{(001)}$. However, the $(100)_{\rm M}$-twin relationship is probably still strongly energetically preferred over more general relationships, i.e., it should correspond to a local minimum of the surface energy.

For general orientations, the surface energy $\Phi$ depends on the mutual rotation between the lattices, but this relation cannot be easily expressed. As discussed e.g. by \cite{Bulatov_Acta_2014,Meier_Cambridge_2014}, $\Phi$ roughly corresponds to the degree of coincidence (so called inverse-$\Sigma$), having pronounced minima for low-$\Sigma$ interfaces, but there are additional minima specific for each crystal structure and each material. The high degrees of coincidence are typically reached for a symmetric tilt between the lattices, i.e., when the interface is a plane of mirror symmetry between the lattices, and especially when the tilt axis is a low-index lattice vector.

The energy of a STGB depends on a single scalar parameter, which is the tilt angle $\tau$. The local minima on the $\Phi(\tau)$ function are typically sharp, and the function itself between them is concave, which means that the exact low-$\Sigma$ boundaries are always strongly energetically preferred over interfaces slightly inclined from the exact low-$\Sigma$ orientation. It can be easily seen that all planar interfaces that are of interest for the plastic forming of NiTi are STGBs, in particular the $(001)_{\rm M}$ and $(100)_{\rm M}$ twin interfaces and $(20\bar{1})_{\rm M}$ kwink interfaces in martensite, as well as the $\{41\bar{1}\}$ twins that appear in austenite after the reverse transition.

For B2 (and NiTi austenite in particular), the dependence of the surface energy density on the tilt angle was calculated by \cite{Yadzandoost_CompMaterSci_2017}, by means of the density functional theory. A symmetric tilt between the lattices was assumed, with the rotation axis being parallel with the $[110]_{\rm P}$ lattice vector. The result is qualitatively depicted in Figure \ref{fig_energy}(b). Alongside with the expected minima for low-$\Sigma$ interfaces, such as the $(112)_{\rm P}$-twins, an additional pronounced minumum appeared for the $(114)_{\rm P}$-twin relationship, which agrees well with the fact that such twins stay stable in the austenite phase, when inherited from $(20\bar{1})_{\rm M}$-twins during the reverse transition. Nevertheless, the scatter in the exact orientations of the $(114)_{\rm P}$-twins mentioned in \citep{Molnarova_2020} further confirms that the surface energy is probably not the leading therm for creating these interfaces.

No such calculation has been reported so far for the B19$^\prime$ monoclinic lattice. However, each low-$\Sigma$ (i.e., high coincidence) interface in the B2 structure corresponds, through lattice correspondence, to some low-$\Sigma$ twin or kwink interface between Variant 1 and Variant 2. Hence, the function $\Phi(\tau)$ for  interfaces between two variants of martensite may be expected to adopt the same structure of local minima (Figure \ref{fig_energy}(c)), with one of those most pronounced representing the $(20\bar{1})_{\rm M}$ kwink. {{} Notice that while for B2 austenite, the minimum at $\tau=180^\circ$ represents identity, and thus, zero surface energy, the corresponding minimum for B19$^\prime$ represents a $(100)_{\rm M}$ twin, for which the energy of $\Phi^{(100)}\approx70$~MJ.m$^{-2}$ has been calculated by \cite{Ezaz_Acta_2011}. Still, this minumum can be considered as much deeper than the one for the $(20\bar{1})_{\rm M}$ twin relation, for which the corresponding $(114)_{\rm P}$ minimum in austenite has $\Phi^{(114)}\approx{}320$~MJ.m$^{-2}$ \citep{Yadzandoost_CompMaterSci_2017}.}

On the other hand, the monoclinicity of the B19$^\prime$ lattice precludes existence of low-$\Sigma$ STGBs in a single variant (Figure \ref{fig_energy}(d)). As known from the experiments (see the Supplementary material for \citep{Molnarova_2020}), tilt interfaces in a single variant may exist, but there is probably no preferred orientation, except of the identity ($\tau=180^\circ$). 

\bigskip
As usual for modelling of the martensitic microstructures, we assume that the observed patterns result from a balance between the bulk energy and the surface energy. Because of the lack of quantitative input, especially in the terms of the surface energy, we understand this balance qualitatively only: we expect the microstructure to be composed of bands of finite width, as dictated by finite $\Phi$, and we expect the interfaces to have a tendency to get oriented along the energetically preferred directions, although this tendency can be in most cases outweighted by the bulk energy minimization, that is, by kinematic compatibility requirements. It is very plausible that the scaling of the microstructure as well as the misorientation of the interfaces from the energetically preferred directions may follow from some general equipartitioning rule between the bulk and the surface energy (cf. \citep{JMB1,Seiner_JMPS_2021}). The patterns in plastically formed NiTi are, however, too complex to allow any single expression of the equipartitioning. In the following applications of the model, we do not discuss the particular length-scales at which the microstructure appears.

\section{Modelling results}

In this section, we present explicit calculations supporting the heuristic arguments we brought in Section 2.3.  We take our motivation from the non-linear elastic theory of martensite \citep{JMB1,JMB2}, where the twinning planes and twinning systems are not put into the models \emph{a priori}, but arise from energy minimization, often through a  compatibility analysis. In the same sense, we utilize here the above introduced model, combining the non-linear elastic theory with anisotropic crystal plasticity, and are aiming at rationalizing the experimental observation of $(20\bar{1})_{\rm M}$ plastic twins in terms of energy minimization. And, indeed, the model predicts that plastic   forming can be achieved by introducing plastic twin-like bands, oriented approximately along the $(20\bar{1})_{\rm M}$ plane and constituting approximately the $(20\bar{1})_{\rm M}$ twin relation. In addition, we shown that with activation of the $[100](001)_{\rm M}$ plastic slip, the experimentally observed V-shaped patterns are low-energy microstructures that enable tensile strain accommodation in directions perpendicular to $(001)_{\rm M}$ plane, as indicated by the experiments.

\subsection{Deformation bands at the nucleation stage}
In the first calculation, we analyze the compatible crossing between the $(100)_{\rm M}$ and $(001)_{\rm M}$ twin bands, seen in the micrograph in Figure \ref{pasypopisy}(a) and outlined heuristically in Figures \ref{pasypopisy}(b-g) (see paragraph 2.3.1 for a detailed discussion).  We consider a $(001)_{\rm M}$-compound laminate loaded in tension in the direction perpendicular to the $(001)_{\rm M}$ planes. We assume that there is enough driving force to create small nuclei of $(100)_{\rm M}$ twins in both variants forming the laminate, and that the material then utilizes the $[100](001)_{\rm M}$ dislocation slip to enable the $(100)_{\rm M}$ twin bands to penetrate the $(001)_{\rm M}$ twin bands.

\begin{figure}
 \centering
 \includegraphics[width=0.85\textwidth]{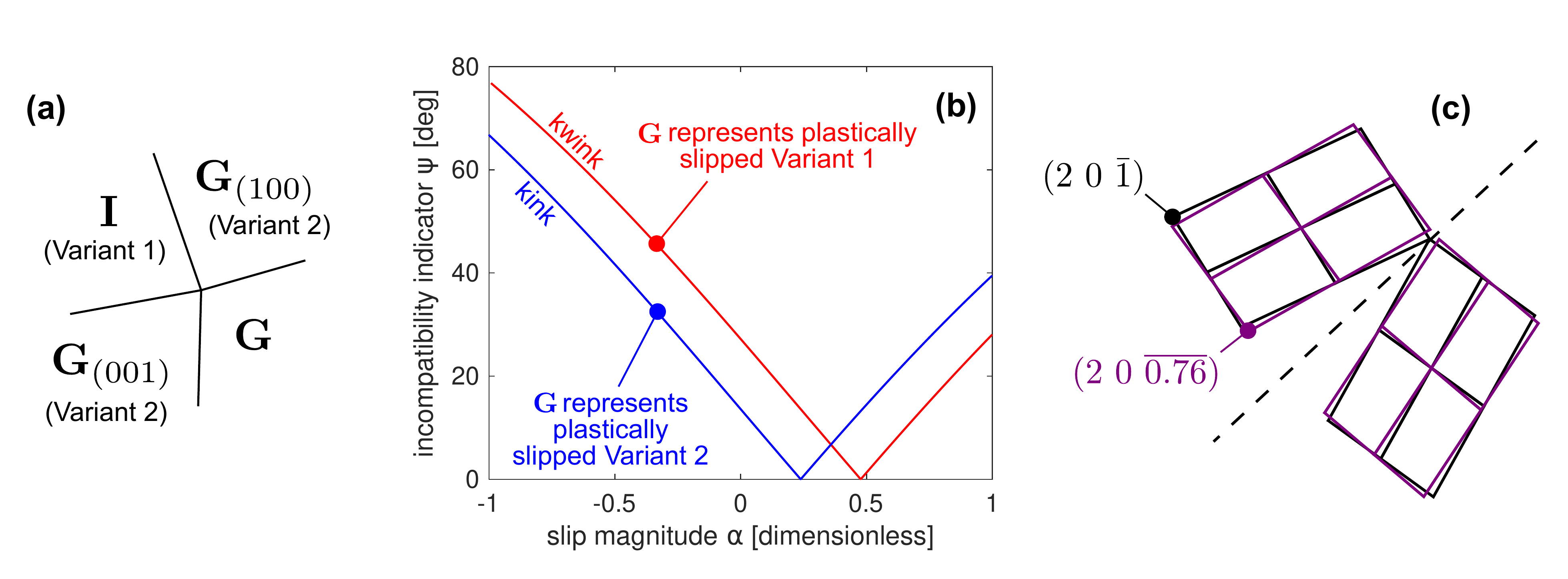}
 \caption{{}(a) the denotation of the deformation gradients for the crossing-twins microstructure; (b) the incompatibility strength (measured by $\psi$) at the interface between ${\mathbf G}$ and ${\mathbf G}_{(001)}$, plotted with respect to the slip magnitude for kwinking (red) and kinking (blue); (c) the difference between the elastic-energy-minimizing kwink (violet, $(2\,\,0\,\,\overline{0.76})_{\rm M}$-twin relation) and the exact $(20\bar{1})_{\rm M}$ plastic twin (black). } \label{penetration}
\end{figure}

The continuum-mechanics model of the penetration of the bands is drawn in Figure \ref{penetration}(a). The matrix (deformation gradient ${\mathbf I}$, representing Variant 1) holds a $(100)_{\rm M}$-twin relation with Variant 2 represented by the deformation gradient ${\mathbf G}_{(100)}$, and a $(001)_{\rm M}$-twin relation with Variant 2 represented by the deformation gradient ${\mathbf G}_{(001)}$, where ${\mathbf G}_{(100)}$ and ${\mathbf G}_{(001)}$ differ just by a rotation. Explicitly, ${\mathbf G}_{(001)}={\mathbf F}$, and
\begin{equation}
 {\mathbf G}_{(100)}={\mathbf R}_{\left(\theta=13.6^\circ\right)}{\mathbf F}=\left(\begin{array}{cc}  0.9720 & -0.0033\\  0.2351 & 1.0280 \end{array} \right).
\end{equation}

We do not postulate which variant is represented by the fourth deformation gradient ${\mathbf G}$,  or what are the orientations of the interfaces separating the region with this deformation gradient from the $(100)_{\rm M}$ and $(001)_{\rm M}$ bands, but we require ${\mathbf G}$ to satisfy the compatibility conditions. As shown by  \cite{BD1,BD2}, in the two-dimensional setting there always exists a compatible connection between two deformation gradients, if these two gradients have the same determinant. If ${\mathbf G}$ is given by (\ref{GRFP}), it represents just volume-preserving operations on the lattice, and thus, the condition on the determinant is satisfied.
Then, for any suggested ${\mathbf G}^0{\mathbf P}(\alpha)$, where ${\mathbf G}^0\in{}\left\{{\mathbf I},{\mathbf F}\right\}$ represents one variant of martensite and ${\mathbf P}(\alpha)$ represents a $[100](001)_{\rm M}$ slip with arbitrary magnitude $\alpha$,  there always exist rotation matrices ${\mathbf R}_{(001)}$ and ${\mathbf R}_{(100)}$ and corresponding vectors ${\mathbf b}_{(001)}$, ${\mathbf b}_{(100)}$, ${\mathbf m}_{(001)}$, and ${\mathbf m}_{(100)}$ such that
\begin{equation}
 {\mathbf R}_{(100)}{\mathbf G}^0{\mathbf P}(\alpha)-{\mathbf G}_{(100)}={\mathbf b}_{(100)}\otimes{\mathbf m}_{(100)}
\end{equation}
and
\begin{equation}
 {\mathbf R}_{(001)}{\mathbf G}^0{\mathbf P}(\alpha)-{\mathbf G}_{(001)}={\mathbf b}_{(001)}\otimes{\mathbf m}_{(001)}. \label{eq_G001}
\end{equation}

The compatibility at the crossing point is then achieved if and only if ${\mathbf R}_{(100)}={\mathbf R}_{(001)}$; the sought gradient is then ${\mathbf G}={\mathbf R}_{(100)}{\mathbf G}^0{\mathbf P}(\alpha)={\mathbf R}_{(001)}{\mathbf G}^0{\mathbf P}(\alpha)$. The deviation from the compatibility can be measured by a scalar parameter called \emph {the incompatibility indicator} \citep{JMPS,Stupkiewicz_CMT_2012}, which is 
\begin{equation}
 \psi=\arccos{\left(\frac{{\rm tr}{({\mathbf R}_{(100)}{\mathbf R}^{\rm T}_{(001)})}-1}{2}\right)},\label{incompatibility}
\end{equation}
and which equals zero for perfect compatibility.  Three disjoint cases are now to be discussed:
\begin{enumerate}
 \item{} ${\mathbf G}^0={\mathbf I}$, $\alpha=0$, which means that ${\mathbf G}$ represents Variant 1 without plastic slip. The rotations ${\mathbf R}_{(100)}$ and ${\mathbf R}_{(001)}$ then represent the $(001)_{\rm M}$-and $(100)_{\rm M}$-twin relations, respectively, that the Variant 1 can hold to Variant 2. The former is the situation visualized in Figure \ref{pasypopisy}(b). As seen already from the visualization, there is a strong incompatibility in this case, with $\psi=27.2^\circ$. 
 \item{} ${\mathbf G}^0={\mathbf I}$, $\alpha\neq{}0$, which is the case visualized in Figure \ref{pasypopisy}(e). Here, a kwink interface appears between  the regions with ${\mathbf G}_{(001)}$ and ${\mathbf G}$. The value of the incompatibility indicator $\psi$ depends on the magnitude of plastic slip $\alpha$ as plotted in Figure \ref{penetration}(b). The perfect compatibility is achieved for \begin{equation}
      \alpha=2\gamma=0.4770,                                                                                                                                                                                                                                                                                                                                                                                        
                                                                                                                                                                                                                                        \end{equation}
which can be also deduced from the mirror symmetries between the latices in ${\mathbf I}$ and ${\mathbf G}_{(001)}$, and in  ${\mathbf I}$ and ${\mathbf G}_{(100)}$.  Importantly, the kwinking plane corresponding to this value of $\alpha$ is $(2\,\,0\,\,\overline{0.76})_{\rm M}$, which differs from $(20\bar{1})_{\rm M}$ only by 4.26$^\circ$. Hence, the lattices neighboring over this kwink interface would hold a $(2\,\,0\,\,\overline{0.76})_{\rm M}$-twin relation, which differs from the $(20\bar{1})_{\rm M}$-twin relation by a small rotation visualized in Figure \ref{penetration}(c). In reality, the deviation from the $(20\bar{1})_{\rm M}$-twin relation can be even much smaller, as the interface may tend to reduce its surface energy by tilting into a low-$\Sigma$ orientation. Also, as discussed in \citep{Mohammed_Acta_2020,Karki_Acta_2020}, irrational twinning planes inclined just few degrees from lattice planes may tend to have a piece-wise rational orientation with steps having the meaning of twinning dislocations. As a result, in high-resolution TEM observations, the interface might be locally in a perfect alignment with the closest low-$\Sigma$ plane, which is $(20\bar{1})_{\rm M}$. For both these reasons, the deviation seen in Figure \ref{penetration}(c) should be understood as rather a maximum deviation that one can expect at the kwink interface that forms due to compatible crossing between $(100)_{\rm M}$ and  $(001)_{\rm M}$ twin bands.
\item{} ${\mathbf G}^0={\mathbf F}$, $\alpha\neq{}0$, which means that there is only one variant inside of the $(100)_{\rm M}$ twin band, and the incompatibility is compensated purely by the dislocation slip. There exists a solution that enables full compatibility, which is $\alpha=\gamma$ (the dependence of the incompatibility indicator on $\alpha$ is plotted in Figure \ref{penetration}(b)). This solution would represents a kink. Because the kwink from the first solution and the kink from this second solution carry the same deformation gradient
\begin{equation}
 {\mathbf F}{\mathbf P}(\gamma)={\mathbf P}(2\gamma), \label{equivPF}
\end{equation}
the compatibility equation (\ref{eq_G001}) between ${\mathbf G}$ and ${\mathbf G}_{(001)}$ is satisfied for the same ${\mathbf m}_{(001)}$, which means the kink runs again along the $(2\,\,0\,\,\overline{0.76})_{\rm M}$ plane of the matrix.

Regardless of the fact that such interfaces were indeed observed occasionally in the experiment (see a perfect $(20\bar{1})$-oriented kink band in Figure 7 of \citep{Molnarova_2020}), the second solution requires some more detailed discussion. Most importantly, the relation (\ref{eq_G001}) means that these two solutions are indistinguishable from the point of view of continuum mechanics. That is, based just on the compatibility conditions, we cannot predict which of the solutions prevails. Moreover, the gradient ${\mathbf F}{\mathbf P}(\alpha)$ does not distinguish between the case when the plastic strain results from motion of individual dislocations with full $[1\,0\,0]_{\rm M}$ Burgers vectors, as usual for plasticity, and the case when the $(001)_{\rm M}$ planes slip along each other homogeneously, and each unit cell deforms as undergoing a slip with a partial (irrational) Burgers vector, which might be understood as a $(001)_{\rm M}$-twinning dislocation. From this point of view, we realize that the first solution (which is Variant 1 slipped homogeneously with amplitude $2\gamma$) can represent exactly the same lattice as the second solution (which is  Variant 2 slipped homogeneously with amplitude $\gamma$) at least in the terms of the shape of the unit cells, as commented below. Hence, while the compatibility calculation can \emph{explain} the strip-like contrast inside of $(100)_{\rm M}$-twin bands in Figure \ref{pasypopisy}(a) (because the kwinks are the solution to the compatibility problem), it cannot \emph{predict it} (because there also is a kink-related solution). Here we probably need to utilize the extension of the model by the surface energy term $W_{\rm surf}$, which is reduced by high coincidence for the kwink, but not for the kink. At the nucleation stage, where the newly appearing bands are very narrow, the surface energy may be decisive.  
\end{enumerate}

\begin{figure}[t]
 \centering
 \includegraphics[width=0.65\textwidth]{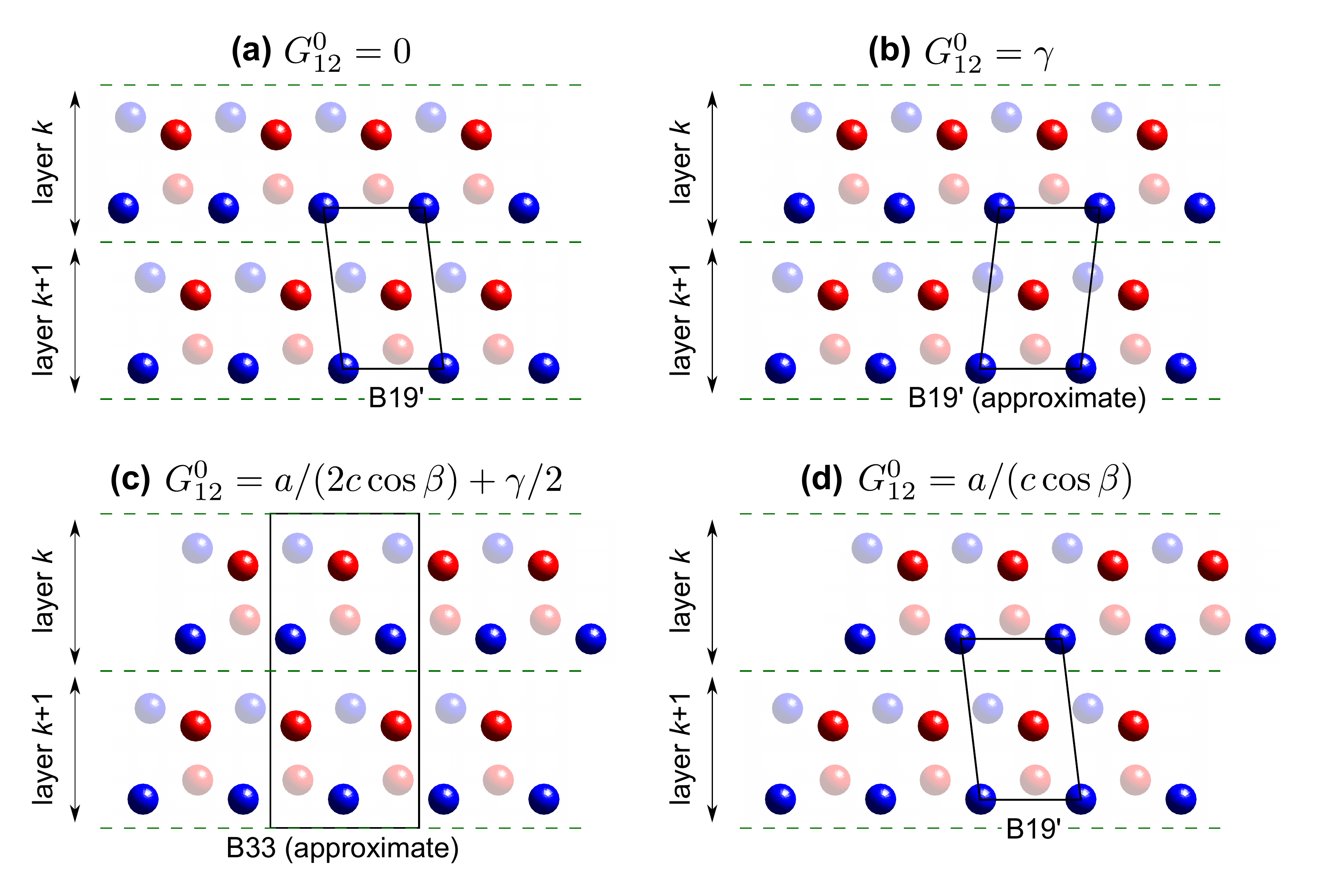}
 \caption{Atomistic scale representation of NiTi martensite mechanics as mechanics of a layered medium. The $k-$th layer slips along the $(k+1)-$th with a magnitude that, at the continuum level, would be associated with the deformation gradient (\ref{G012}). The gradual increase in $G^0_{12}$ leads to the following subsequent low-energy stages: (a) the initial lattice; (b) reoriented martensite within the same EPN; (c) a thought orthorhombic B33 structure, considered as a low-energy state between two EPNs \citep{Gao,pathways}; (d) the initial lattice after a passage of a full $[100](001)_{\rm M}$ dislocation. The lattices in (b) and (c) are approximate, differing from the perfect lattice by small rearrangements of internal atoms inside the cell, i.e., the slipping must be accompanied with a low-amplitude shuffle. The (a)$\rightarrow$(b) path can represent either $(001)_{\rm M}$ twinning or $(100)_{\rm M}$ twinning, the (a)$\rightarrow$(d) path represents either a full dislocation slip or a sequence of $(001)_{\rm M}$ and $(20\bar{1})_{\rm M}$ twinnings. }\label{layers}
\end{figure}

The discussion of the third case above illustrates how deep is the coupling between the slip and twinning in the considered $(001)_{\rm M}$ plane of B19$^\prime$ at the finest scale. If a deformation gradient can be, up to a rotation, expressed as 
\begin{equation}
 {\mathbf G^0}=\left(\begin{array}{cc}
                      1 & G^0_{12}\\ 0 & 1
                     \end{array} \right), \label{G012}
\end{equation}
where $G^0_{12}$ is some scalar parameter, then ${\mathbf G^0}$ may always represent either Variant 1 or Variant 2 with different magnitudes of slip. For specific slip magnitudes ($G^0_{12}=na/(c\cos\beta)\pm\gamma$, where $n\in{}\mathbb{Z}$, and the sign depends on which variant we take as the parent one) the deformation gradient can be applied homogeneously to all unit cells (\emph{via} an inverse-sense Cauchy-Born rule), resulting in mapping one of the variants onto another. This leads to a layer-like interpretation of the atomistic-scale mechanics of NiTi, as outlined in Figure \ref{layers}, and as discussed often in the literature concerned with first-principles modelling of NiTi (e.g. \citep{Ezaz_Acta_2011}); see also for example \citep{Krcmar_PRM_2020}, where even the austenite$\leftrightarrow$martensite transition is interpreted as formation of faults in stacking of the $(001)_{\rm M}$-parallel layers.
The same shear deformation of the unit cells can be equivalently carried by a full dislocation ((a)$\rightarrow$(d) in Figure \ref{layers}), resulting in plastic strain without martensite reorientation, or by a sequence of twinning dislocations that sum up into the same slip magnitude (the sequence (a)$\rightarrow$(b)$\rightarrow$(d) in Figure \ref{layers}).

 We formulate our model at much longer length-scales, where we do not need to distinguish between these cases, or can distinguish between them based on experimental observations. Nevertheless, these atomistic-scale considerations indicate that the B19$^\prime$ structure with highly anisotropic plastic slip is indeed very similar to layered media in its behavior, or even to the physical model of pile of papers studied in \citep{Wadee}, which brings an additional justification of the approach presented in this paper. 
Predictions of the partitioning of $G^0_{12}$ between the  $(001)_{\rm M}$ twinning and $[100](001)_{\rm M}$ dislocation slip fall beyond the scope of this paper and beyond the chosen continuum-level description. Nevertheless, from the analyses in the following two subsections, it is obvious that the model requires deformation gradients ${\mathbf P}(\alpha)$ with arbitrary and continuously changing $\alpha\in{\mathbb R}$, for enabling a homogeneous plastic deformation of the polycrystal and for capturing fully the observed pattern formation. Such deformation gradients cannot be visualized or interpreted at the atomistic scale, they must arise from averaging of shear strains from localized dislocation slip over larger regions. Thus, the chosen continuum-level crystal plasticity framework appears to be a properly chosen tool for describing this behavior.

\subsection{Tension-induced plastic deformation of oriented martensite}

As discussed in Section 2.3.3. the plastic twinning may serve as a compensation of perpendicular shears arising due to the $[100](001)_{\rm M}$ dislocation slip, when fully oriented martensite is loaded in tension along the $[10\bar{1}]_{\rm M}$ direction. In principle, the compensation can be achieved by any shearing mechanism that comprises elongation along the  $[10\bar{1}]_{\rm M}$ direction and, simultaneously, perpendicular shears with an opposite sign than the  $[100](001)_{\rm M}$ dislocation slip. We consider that this shearing mechanism has the form of deformation bands (such as twins, kinks or kwinks) that occupy some volume fraction $\lambda$ of the grain. The optimum is reached when the considered mechanism can do the compensation with the smallest $\lambda$, and, at the same time maximally contributes to the axial elongation.

From the point of view of the continuum-mechanics model, the resulting microstructure is a laminate that consists of Variant 1 (i.e. the matrix) plastically deformed with the $[100](001)_{\rm M}$ slip magnitude $\alpha_1$, and bands of heavily slipped Variant 2 that forms kwinks in the matrix. The plastic slip inside of the band is some $\alpha_2$, and thus, the total deformation gradient is ${\mathbf F}{\mathbf P}(\alpha_2)$ up to an orthogonal rotation ${\mathbf R}$ that ensures the compatible connection between the matrix and the bands, i.e.,
\begin{equation}
 {\mathbf R}{\mathbf F}{\mathbf P}(\alpha_2)-{\mathbf P}(\alpha_1)={\mathbf a}\otimes{}{\mathbf n}
\end{equation}
for some shearing vector ${\mathbf a}$ and some kwink-interface normal ${\mathbf n}$. Since ${\mathbf F}$ and ${\mathbf P}$ commute for any magnitude of the slip, and so do the matrices ${\mathbf P}(\alpha_1)$ and ${\mathbf P}(\alpha_2)$, the same deformation gradient is achieved regardless of what is the sequence of strains the material inside the bands undergoes. Either it can slip first with the same magnitude as the matrix ($\alpha_1$) and then undergo additional slip with magnitude ($\Delta\alpha=\alpha_2-\alpha_1$) and martensite reorientation such that the kwink bands are formed, or it can create the bands with the deformation gradients ${\mathbf F}{\mathbf P}(\Delta\alpha)$ first, and then the matrix and the bands can both undergo dislocation slip with the magnitude $\alpha_1$. This confirms that the proposed mechanism satisfies our requirement that both processes must occur simultaneously, with each increment in the plastic slip of the matrix being compensated by a small growth of the kwink bands, and vice versa.    

The total deformation gradient of the laminate is then
\begin{equation}
{\mathbf G}(\alpha_1,\alpha_2,\lambda,\theta)={\mathbf Q}(\theta)\left[\lambda{\mathbf R}{\mathbf F}{\mathbf P}(\alpha_2)+ (1-\lambda){\mathbf P}(\alpha_1)\right]={\mathbf Q}(\theta)\left[\lambda{\mathbf R}{\mathbf F}{\mathbf P}(\Delta\alpha)+ (1-\lambda){\mathbf I}\right]{\mathbf P}(\alpha_1), \label{Glamin}
\end{equation}
where ${\mathbf Q}$ is a rotation matrix that compensates possible artificial rotations of the grain from the tension axis, and $\theta$ is the angle of the rotation in the sense of definition (\ref{rotace}). Notice that the twin-like relationship over the kwink interface is determined solely by $\Delta\alpha$, i.e., by the jump of the slip magnitude over the interface, and is, thus, independent on the total slip in the matrix.

For a perfect compensation, we require
\begin{equation}
{\mathbf G}(\alpha_1,\alpha_2,\lambda,\theta) = {\mathbf G}_{[10\bar{1}]}(\varepsilon_{[10\bar{1}]}), \label{natazeni}
\end{equation}
where ${\mathbf G}_{[10\bar{1}]}(\varepsilon_{[10\bar{1}]})$ is a deformation gradient representing the elongation of the grain along the tension direction with the axial strain of $\varepsilon_{[10\bar{1}]}$ and proportional shrinking in the perpendicular direction, such that the area of the grain in the $(010)_{\rm M}$ plane remains constant. This assumption of volume conservation arises from the fact that we want to reach ${\mathbf G}_{[10\bar{1}]}(\varepsilon_{[10\bar{1}]})$ by combining strictly volume-preserving mechanisms, which are the slip and the martensite reorientation. This is obviously a simplification, since the plastic forming of the grain is expected to relax elastic strains, where the shrinking in the cross-section is dictated by the Poisson's ratio. However, as explained in Section 3.3, we do not attempt to involve the energy from the external loads directly in the model. Instead, we search for mechanisms with low or zero $W_{\rm bulk}+W_{\rm surf}$ such that the deformation of the material complies with the external loads.

\begin{figure}
 \centering
 \includegraphics[width=\textwidth]{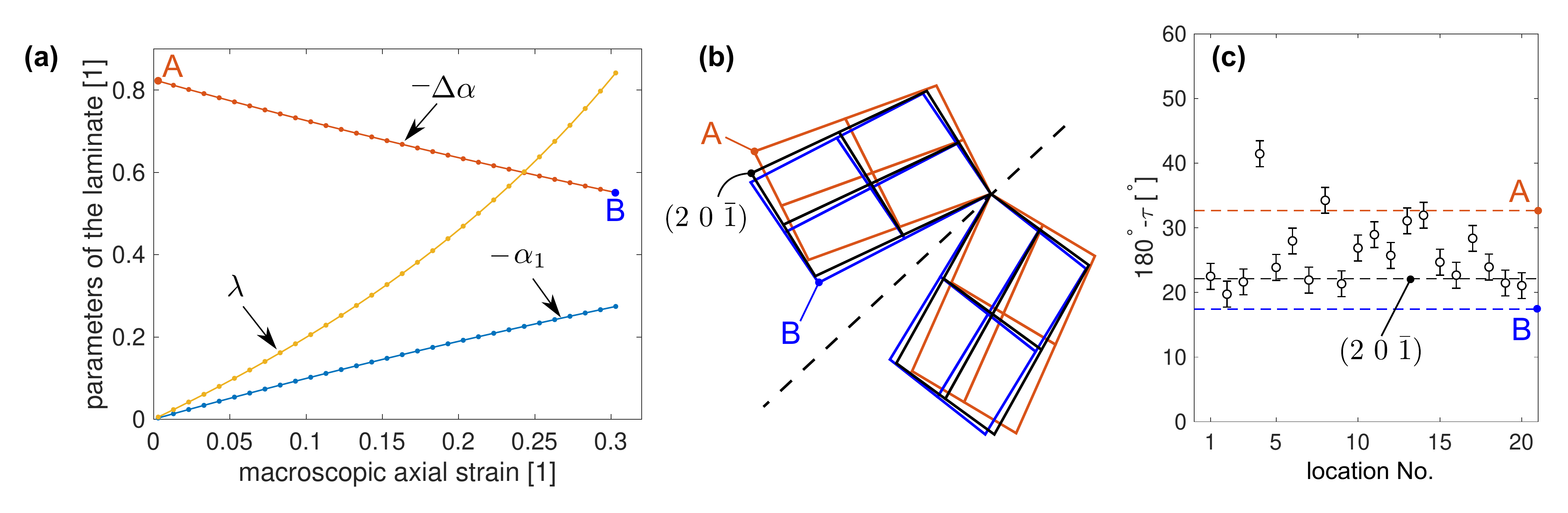}
 \caption{(a) parameters of the construction (\ref{Glamin}) resulting in the closest approximation of pure axial extension of the of the grain in sense of the minimization (\ref{minimization}). The decisive factor for the orientation of kwink is the jump of the slip magnitude $\Delta{}\alpha$. The other parameters are the volume fraction of the kwink bands $\lambda$ and slip magnitude in the matrix $\alpha_1$. The rotation $\theta$ (not shown) is negligibly small. (b) visualizations of twin-like relationships over kwinking planes for two limiting cases. The black lattices hold a perfect $(20\bar{1})$-twin relationship. (c) an experimental dataset from \citep{Sittner_HRTEM_SSR}, showing measured tilt angles $\tau$ for 20 different locations in a fine-grained textured polycrystal. The dashed lines marked A and B correspond to the limiting cases from (a), the middle dashed line to a perfect $(20\bar{1})$-twin relationship. }
 \label{orientation}
\end{figure}

The condition (\ref{natazeni}) has to be solved numerically. By minimizing 
\begin{equation}
 |{\mathbf G}(\alpha_1,\Delta\alpha,\lambda,\theta)-{\mathbf G}_{[10\bar{1}]}(\varepsilon_{[10\bar{1}]})|{\rightarrow}{\rm min}_{{\alpha_1,\Delta\alpha,\lambda,\theta}}, \label{minimization}
\end{equation}
where $|\cdot|$ is the Frobenius norm,
we can achieve a sequence of parameters of the microstructure such that the resulting laminate follows the prescribed strains in an optimal way. For $\varepsilon_{[10\bar{1}]}$ up to 30\%\footnote{Notice that the numerical minimization in (\ref{minimization}) becomes ill-posed for $\varepsilon_{[10\bar{1}]}\rightarrow{}0$, because with $\lambda\rightarrow{0}$ the minimum is reached for arbitrary $\Delta\alpha$ and arbitrary $\theta$. For this reason, the numerical minimization was done starting from $\varepsilon_{[10\bar{1}]}=3\times10^{-3}$ instead of from zero.}, the result of the numerical minimization is shown in Figure \ref{orientation}(a). As expected, with increasing $\varepsilon_{[10\bar{1}]}$ also the slip in the matrix $\alpha_1$ and the volume fraction of the kwink bands $\lambda$ increase in their magnitudes, starting from 0 for $\varepsilon_{[10\bar{1}]}\rightarrow{}0$ (notice that due to geometry reasons, both $\alpha_1$ and $\Delta\alpha$ are negative). In contrast, the jump of the slip magnitude over the kwink interface, $\Delta\alpha$, remains between $\Delta\alpha=-0.82$ (point A in Figure \ref{orientation}(a)) and $\Delta\alpha=-0.55$ (point B in Figure \ref{orientation}(a)) for the whole range of axial strains. Both these limiting values correspond to kwinking planes quite close to the $(20\bar{1})_{\rm M}$ orientation, and, consequently, the twin relations over the the kwink interface are close to  $(20\bar{1})_{\rm M}$-twinning. This is visualized in Figure \ref{orientation}(b). The deviations of the kwink plane orientation from $(20\bar{1})_{\rm M}$ are 5.21$^\circ$ for the point A and $-$2.18$^\circ$ for the point B; the perfect $(20\bar{1})_{\rm M}$-twin relationship is reached for approximately $\varepsilon_{[10\bar{1}]}=22.9\,\%$. Nevertheless, similarly as for the case of the crossing-twins microstructure discussed in the previous subsection, the exact low-$\Sigma$ can be energetically preferred for the whole range of $\varepsilon_{[10\bar{1}]}$. The rotation angle $\theta$ (not shown) did not reach over 0.1$^\circ$ for any $\varepsilon_{[10\bar{1}]}$, i.e., which means there is no artificial additional rotation resulting from the construction of the deformation gradient ${\mathbf G}$ introduced above.   

Figure \ref{orientation}(c) shows experimentally determined tilt angles $\tau$ extracted from high-resolution TEM images (see the Supplementary material to \citep{Sittner_HRTEM_SSR} for more details on the experiment) of kwink interfaces in a wire deformed to 15\% of axial strain. Except for one or two outliers, all experimental values lie between the limiting case A (zero plastic strain) and the perfect $(20\bar{1})_{\rm M}$-twin orientation, or overlap with this interval by their error bars. There are many points lying very close to $(180^\circ-\tau)=22.12^\circ$, while the mean value is shifted to higher $(180^\circ-\tau)$ angles. That is consistent with the model prediction that for $\varepsilon_{[10\bar{1}]}<23\,\%$ the tilt angle should be closer to the limiting state A, but there is a tendency to stick to the low$-\Sigma$ orientation at least in some segments of the interface. However, considering the heterogeneous distribution of the strains across individual grains and the experimental scatter, any quantitative conclusions are hard to draw.  

\subsection{Formation of V-shaped patterns}

The last benchmark test for the model is the formation of the experimentally observed V-shaped patterns, consisting always of one $(100)_{\rm M}$ twin band and one $(20\bar{1})_{\rm M}$ kwink band. As discussed heuristically in subsection 2.3.4, such patterns can accommodate tensile strains in direction perpendicular to the $(001)_{\rm M}$ planes, which means they may contribute to stress relaxation in grains loaded in such direction. Here we need to show that these patterns are energetically not expensive,  disclinations arising at the triple points can be fully compensated be a mechanism invisible for the electron microscopy, which is the coordinated $[100](001)_{\rm M}$ plastic slip, where the midrib plane (between lattices $A$ and $B$ in Figure \ref{wedge}) serves as a source or sink for the additional dislocation cores. We want to enumerate the amplitude of plastic slip enabling a full compensation of the disclination, and to show that in the deformed configuration the pattern is indeed a perfect V-shape, that is, the interfaces (7)-to-(6) and (5)-to-(3) in Figure \ref{cartoon} become exactly parallel to each other. For simplicity, we assume that the kwink band is exactly $(20\bar{1})$ oriented, which means that the deformation gradient in this band is exactly equal (up to a rotation) to ${\mathbf F}^\prime$.

\begin{figure}
 \centering
 \includegraphics[width=0.7\textwidth]{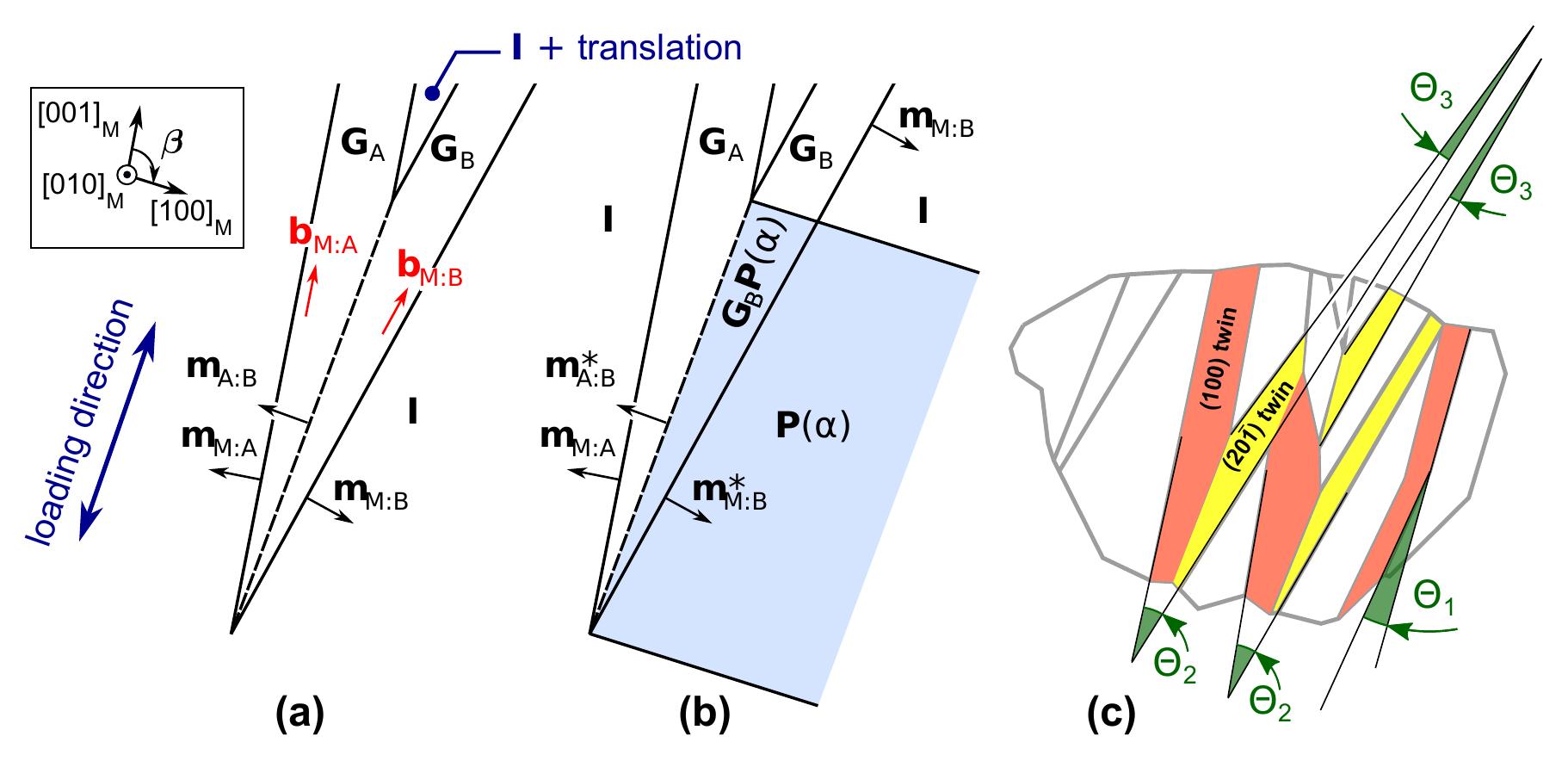}
 \caption{The notation needed for the compatibility analysis for the V-shaped microstructure. (a) a theoretical construction from three homogeneous deformation gradient; (b) a compensation of the disclination dipole through a plastic slip in the shaded region; (e) contours of twin interfaces extracted from the micrograph in Figure \ref{wedge}(a). The angles $\Theta_i$ measuring the deviation of the observed pattern from the idealized case shown in (a).}
 \label{kompa}
\end{figure}

The notation used for the compatibility analysis is outlined in Figure \ref{kompa}(a): we assume regions of homogeneous deformation gradients ${\mathbf G}_A$ and ${\mathbf G}_B$ surrounded by the matrix ($M$, deformation gradient ${\mathbf I}$). Between these homogeneous deformation gradients, we assume planar interfaces perpendicular (in the reference configuration) to unit vectors ${\mathbf m}_{M:A}$, ${\mathbf m}_{M:B}$ and ${\mathbf m}_{A:B}$.

As the lattices $M$ and $A$ are $(100)_{\rm M}$-twin related, the deformation gradient representing $A$ must be 
\begin{equation}
 {\mathbf G}_A={\mathbf R}_{\left(\theta=13.6^\circ\right)}{\mathbf F}=\left(\begin{array}{cc}  0.9720 & -0.0033\\  0.2351 & 1.0280 \end{array} \right). \label{GA_Vshape}
\end{equation}
Similarly, the $(20\bar{1})$-twin relation between $B$ and $M$ implies that 
\begin{equation}
 {\mathbf G}_B={\mathbf R}_{\left(\theta=-22.12^\circ\right)}{\mathbf F}^\prime=\left(\begin{array}{cc} 0.9264 & 0.0144 \\ -0.3766 & 1.0736 \end{array} \right).
\end{equation}

The shears contained in the lattices $A$ and $B$ are represented by the shearing vectors ${\rm b}_{M:A}$ and ${\rm b}_{M:B}$, both shifting the region above the V-shaped microstructure approximately along the loading direction. As the result, this region remains having the lattice orientation of the matrix, but is translated along the loading direction and the grain becomes elongated. {{} Quantitatively, the elongation of the grain along the loading direction (or, more precisely, along its projection onto the considered $(010)_{\rm M}$ plane) is given by the elongation of the material inside of the V-shaped pattern. As the loading direction ${\mathbf p}$ is approximately parallel to the $A$:$B$ interface and approximately perpendicular to the $(001)_{\rm M}$ plane in the matrix, which means ${\mathbf p}\approx(0;1)$,  we can calculate this elongation directly from the deformation gradient (\ref{GA_Vshape}) as the strain 
\begin{equation}
 \varepsilon={\mathbf p}{\mathbf G}_A{\mathbf p}^{\rm T}-1=2.8\,\%
\end{equation}
 times the length of the $A$:$B$ interface in the reference configuration. \cite{Ezaz_MSEA_2020} used digital image correlation (DIC) to measure axial strains inside of the $(100)_{\rm M}$ and $(20\bar{1})_{\rm M}$ bands, obtaining scattered axial-strain maps with $\varepsilon$ ranging between 2~\% and 3.5~\%, which matches very well our continuum-level result of $\varepsilon=2.8$~\%.}

\bigskip
Now, for the compatibility to be achieved for the whole pattern,
there must exist such vectors ${\mathbf b}$ and ${\mathbf m}$ such that the compatibility equation (\ref{compat_G}) between  lattices $A$ and $B$,
\begin{equation}
 {\mathbf G}_A-{\mathbf R}{\mathbf G}_B={\mathbf b}\otimes{\mathbf m}, \label{compat_AB} 
\end{equation}
is satisfied for $\theta=0$ (i.e., ${\mathbf R}={\mathbf I}$). As expected from the heuristic construction in section 2.3, this condition is met only approximately: the solution of (\ref{compat_AB}) gives $\psi = 0.78^\circ$, with the incompatibility indicator $\psi$ defined in the same sense as in (\ref{incompatibility}), which means there is a small misfit between the rotations that must be compensated by additional strains, either elastic, or plastic. Using the terminology of \citep{Inamura,Mullner_ZMk_2006,Mullner_Acta_2010}, there is a disclinaton dipole surrounding the midrib plane of the wedge.

\bigskip
The geometrically simplest construction of the compensation of the disclinations by plastic slip is outlined in Figure \ref{kompa}(b). It is the same mechanism as we proposed in Figure \ref{cartoon}.  Here, a band of plastically slipped material approaches the midrib of the V-shaped microstructure from one side, bringing an additional plastic strain ${\mathbf P}(\alpha)$. If the band runs along the slip plane $(001)_M$, the compatibility between the matrix and ${\mathbf P}(\alpha)$ is trivially satisfied without any additional rotation, and so is also between ${\mathbf G}_B{\mathbf P}(\alpha)$ and ${\mathbf G}_B$ inside of the band. The only change of the geometry is that the midrib plane $A$:$B$ and the interface between ${\mathbf P}(\alpha)$ and ${\mathbf G}_B{\mathbf P}(\alpha)$ should have altered orientations (denoted ${\mathbf m}^*_{A:B}$ and ${\mathbf m}^*_{M:B}$, respectively) in the reference configuration. However, from the compatibility conditions for the $M$:$B$ interfaces we get 
\begin{equation}
\begin{split}
{\mathbf b}_{M:B}\otimes{{\mathbf m}_{M:B}}={\mathbf G}_B-{\mathbf I}=\left[{\mathbf G}_B{\mathbf P}(\alpha)- {\mathbf P}(\alpha)\right]{\mathbf P}^{-1}(\alpha)\\
=({\mathbf b}^*_{M:B}\otimes{\mathbf m}^*_{M:B}){\mathbf P}^{-1}(\alpha)= {\mathbf b}^*_{M:B}\otimes\left[{\mathbf P}^{-{\rm T}}(\alpha){\mathbf m}^*_{M:B}\right],
\end{split} \label{bmbm}
\end{equation}
where ${\mathbf b}^*_{M:B}$ is the shearing vector at the $M$:$B$ interface inside of the plastically strained band. The relation (\ref{bmbm}) directly implies that \begin{equation}
{\mathbf b}_{M:B}\parallel{}{\mathbf b}^*_{M:B}{\, \, \mbox{\, \, and \, \,}\, \,}{\mathbf m}_{M:B}\parallel{\mathbf P}^{-{\rm T}}(\alpha){\mathbf m}^*_{M:B},                                                                                                                                                                                                                                                                                             \end{equation}
from which the second parallelism means that in the deformed configuration the interface orientation between the matrix and the $B$-lattice band is the same both in the plastically strained region and outside of it. In other words, the outer faces of the V-shaped microstructure remain planar even if there is a plastically deformed band approaching the midrib as sketched in Figure \ref{kompa}(b).
Thus, we can conclude that the mechanism in Figure \ref{cartoon} is indeed resulting in a perfect V-shape pattern, despite the fact that the interfaces (6)-to-(7) and (3)-to-(5) are not parallel in the reference configuration.

The condition for the slip magnitude $\alpha$ to compensate the disclinations can be written as
\begin{equation}
 \det\left[{\mathbf G}_A-{\mathbf G}_B{\mathbf P}(\alpha)\right]=0, \label{detcond}
\end{equation}
which results in $\alpha=-0.01505$, i.e., just 1.5\% plastic strain is sufficient. The condition (\ref{detcond}) means that there exists a non-zero vector in the $(010)_{\rm M}$ plane that deforms equivalently under ${\mathbf G}_A$ and under ${\mathbf G}_B{\mathbf P}(\alpha)$, which is the vector lying in the $A$:$B$ interface. Such a  vector exists if and only if there is a zero rotation misfit between ${\mathbf G}_A$ and ${\mathbf G}_B{\mathbf P}(\alpha)$, which implies that $\psi=0$ and full compatibility is achieved. 

The plastically strained band is assumed to be infinite to the right in the construction in Figure \ref{kompa}(b), and thus, there must be some evidence of the plastic slip in this band quite far away from the V-shaped microstructure. When discussing the compatibility of kink patterns in layered media, \cite{Inamura} suggested that this plastically strained band may terminate at another wedge-like microstructure oriented upside-down with respect to the discussed one. However, in our case there are two V-shaped microstructures with the same orientation in the 
discussed grain, so probably the slipped band rather terminates at the grain boundary that may act as a source of dislocations for the slip. Notice that the $(001)_{\rm M}$ planes along which the slip is assumed are unfavorably oriented with respect to the loading direction, having a quite small Schmid factor. This means that the driving force for the slip must come from stress localization at the disclination, i.e., the slip and the formation and growth of the V-shaped microstructure are strongly coupled.   

In the discussed representative grain (Figure \ref{wedge}(a)), an isolated $(100)_{\rm M}$ twin band is seen running between the V-shaped microstructures and the grain boundary. By a closer inspection, we can observe that this band is not straight, but it changes its orientation approximately in the same height where the midribs of the V-shaped microstructures appear. As the $(100)_{\rm M}$ twinning plane is tightly connected to the crystallographic orientation, this may be an indication that the lower half of the band interacts with some homogeneous and localized plastic strain. The change of the orientation of the band is represented by the angle $\Theta_1$ in Figure (\ref{kompa}(c)), which is $\Theta_1\approx{}9^\circ$.  
Similar such sharp changes in orientations of twinning planes quite far away from the midribs of the V-shaped microstructures can be found also in the more complex patterns reported in \citep{Molnarova_2020,Sittner_HRTEM_SSR}. Hence, we can conclude that there are probably bands or other areas in the grain that undergo localized but homogeneous plastic strain, and the strain accommodation may indeed proceed by the mechanism suggested in Figure \ref{cartoon} (and in Figure \ref{kompa}(b)).

There can obviously exist alternative mechanism for achieving the compatibility, for example by an additional localized slip inside the twin bands, or by a difference between plastic slip amplitudes in the matrix surrounding the V-shaped pattern and in the matrix encapsulating it. In the experiment, the former should be visible as a deviation of the angle between the bands from the theoretically predicted value for perfect $(100)_{\rm M}$ and $(20\bar{1})_{\rm M}$ twin interfaces, and the latter as a non-parallelism of the faces of the individual bands (resulting in wedge-like shapes of the bands). The extracted experimental pattern from the chosen representative grain in Figure \ref{kompa}(c) indicate that both these compensation mechanisms may be, to some extent, active in the material. For the tips of the V-shaped patterns, the theoretical angle should be $\Theta_2\approx{}18^\circ$, while the angle measured from the micrograph in Figure \ref{kompa}(c) gives $\Theta_2\approx{}21^\circ$. Some part of this discrepancy can be, however, described to the fact that the kwinks can be inclined from the exact $(20\bar{1})_{\rm M}$ orientation, as discussed in the previous subsection. At the same time, the interfaces encapsulating the $(20\bar{1})_{\rm M}$ twin are clearly inclined with respect to each other, making the angle of $\Theta_3\approx{}4^\circ$ in Figure \ref{kompa}(c). We can conclude that there might be different plastic strain jumps $\Delta{}\alpha$  at the interfaces encapsulating the kwink band, which results in slightly different interface orentations and narrowing of kwink bands. Again, the wedge-like shapes of some bands are clearly seen in the more complex patterns in \citep{Molnarova_2020,Sittner_HRTEM_SSR}
 
\bigskip
Beyond the above discussed simple mechanisms, the material probably relaxes the elastic energy coming from the disclinations by merging the V-shaped microstructures into more complex patterns. Such mechanisms were in detail discussed for plastic kinks in \citep{Inamura}; some of the morphologies from \citep{Molnarova_2020} clearly resemble the low-energy kink configurations from \citep{Inamura}. Also, by locating the tip of the V-shaped microstructure at the grain boundary or close to it, as seen for the representative grain in Figure \ref{wedge}(a), the grain boundary sliding and the dislocation generated at the grain boundary can contribute to the compensation. Nevertheless, the compatibility misfit still appears at the triple junction between $A$, $B$ and $M$. To remove it, the $A$:$B$ tilt interface should grow until it reaches the opposite boundary of the grain, which may lead to filling the grans with parallel bands, as observed in the experiments \citep{Sittner_SMSE_2023_SMST}.

\section{Conclusions}

{{} In this work, we discussed the mechanism of plastic forming of B19$^\prime$ martensite of NiTi, that is, a material exhibiting very strong anisotropy in terms of the dislocation slip, while having the ability to reorient the lattice within the EPN \emph{via} twinning. We have shown that if the $[100](001)_{\rm M}$ plastic slip and the martensite reorientation between two variants sharing the $(010)_{\rm M}$ plane are active at the same time, 
a novel deformation mechanism that combines kink banding and twinning arises, resulting in new types of deformation bands. Because these bands merge properties of kink bands and twin bands, we decided to term them 'kwink bands'. Based on analyzing experimental observations reported in literature, we concluded that the kwink bands play an essential role in the plastic deformation of B19$^\prime$, including the typical formation of V-shaped patterns. This motivated us to formulate a continuum-mechanics model of kwinking, combining the tools of nonlinear elasticity theory of martensite and crystal plasticity. }
 
Using the proposed continuum-mechanics model, we have shown that the plastic forming of NiTi B19$^\prime$ martensite can indeed proceed through formation of the kwink bands, and that the energy-minimizing orientations of these  bands lie close to $(20\bar{1})_{\rm M}$ planes. Based on this result, we claim that the frequently observed  $(20\bar{1})_{\rm M}$ plastic twins do not need to be treated as an additional twinning system, but that they arise from an interplay between martensite reorientation and highly anisotropic plastic slip. 
 
 Through formation of kwink bands and V-shape patterns, the monoclinic lattice of martensite can achieve macroscopically homogeneous plastic strains (at stress levels well below the yield stress of austenite at the given temperature), despite having only one available slip system. This challenges the broadly accepted modelling paradigm that austenite is the phase more prone to dislocation slip and plastic deformation in NiTi. The kwinking in martensite, as a unique mechanism of plastic forming, may have several consequences for the mechanical performance of irreversibly strained  NiTi, such as the grain refinement and hardening, the strong stabilization of martensite, and the appearance of $(41\bar{1})_{\rm P}$ twin bands after the reverse transformation.

\section*{Acknowledgement}

This work has been financially supported by Czech Science Foundation [project No.22-20181S].

%%%%%%%%%%%%%%%%%%%%%%%%%%%%%%%%%%%%%%%%%%%%%%%%%%%%%%%%%%%%%%%%%%%%%%%%%%%%%%%%%%%%%%%%%%%%%%%%%%%%%%%%%%%%%%%%%%%%%%%%%%%%%%%%%%%%%%%%%%%%%%%%%%%%%%%%%%%%%%%%%%%%%%%%%%%%%%%%%%%%%%%%%%%%%%%%%%%%%%%%%%%%%%%%%%%%%%%%%%%%%%%%%%%%%%%%%%%%%%%%%%%%%%%%%%%%

\end{document}